\documentclass[11pt,a4paper]{JHEP3}
\usepackage{amsmath, amssymb}
\usepackage{euscript}
\usepackage{slashed}
\def\be{\begin{eqnarray}}
 \def\ee{\end{eqnarray}}
\def\0{\nonumber}
\def\del{\partial}
\def\ET{\EuScript T}
\def\trace{\rm tr}
\def\arccoth{{\rm arccoth}}
\newcommand{\lp}{\left(}
\newcommand{\arctanh}{{\rm arctanh}}
\newcommand{\rp}{\right)}
\newcommand{\gammaFct}[1]{\Gamma\left( #1 \right)}
\newcommand{\ArcTanh}[1]{\arctanh\left( #1 \right)}
\global\long\def\pochhammer#1#2{\left(#1\right)_{#2}}
\usepackage[latin9]{inputenc}
\usepackage{amsmath}
\usepackage{amsthm}
\usepackage{amssymb}
\usepackage{esint}
\preprint{SISSA/13/2016/FISI\\ZTF-EP-16-01\\{\tt hep-th/1602.07178} }

\title{Massive fermion model in 3d and higher spin currents}

\author{ L.~Bonora$^{a}$, M.~Cvitan$^{b}$, P.~Dominis Prester$^{c}$, B.~Lima de Souza$^{a}$, I.~Smoli\'{c}$^{b}$
\\\textit{${}^{a}$ International School for Advanced Studies (SISSA),\\Via Bonomea 265, 34136 Trieste, Italy, and INFN, Sezione di
Trieste\\}%
\textit{${}^{b}$ Theoretical Physics Division of Particles and Fields, Faculty of Science, University of Zagreb, 
Bijeni\v{c}ka 32, HR-10000 Zagreb, Croatia\\}%
\textit{${}^{c}$ Department of Physics, University of Rijeka,\\
Radmile Matej\v{c}i\'{c} 2, 51000 Rijeka, Croatia\\}%
E-mail: \email{bonora@sissa.it}, \email{mcvitan@phy.hr},  \email{pprester@phy.uniri.hr}, \email{blima@sissa.it}, \email{ismolic@phy.hr}}

\abstract{We analyze the 3d free massive fermion theory coupled to external sources. 
The presence of a mass explicitly breaks parity 
invariance. We calculate two- and three-point functions of a gauge current and the 
energy momentum tensor and, for instance, obtain the well-known result that
in the IR limit (but also in the UV one) we reconstruct the relevant CS action. 
We then couple the model to higher spin currents and explicitly work out the spin 3 case.
In the UV limit we obtain an effective action which was proposed many years ago as 
a possible generalization of spin 3 CS action. In the IR limit we derive a 
different higher spin action. This analysis can evidently be generalized to higher spins. 
We also discuss the conservation and properties of the correlators we obtain 
in the intermediate steps of our derivation. }


\begin{document}
\maketitle

\section{Introduction}
\label{sec:intro}

In the latest years, field theories, and especially conformal field theories, in
3d have become a favorite ground of research. The motivations for this are 
related both to gravity and to condensed
matter, see for instance \cite{Maldacena,Huh} and references therein, based on
AdS/CFT correspondence, where 3d can feature on both sides of duality. 
Also higher spin/CFT correspondence has raised interest on weakly coupled CFT in 3d, \cite{Vasiliev,Maldacena-Zhiboedov}.
In this context many 3d models, disregarded in the past, are being 
reconsidered \cite{giombi,Closset}. 
This paper is devoted to the free massive fermion model in 3d coupled to various
sources. Unlike the free massless fermion, \cite{GMPTWY}, this model has not been 
extensively studied, although examples of research in
this direction exist, see for instance \cite{Babu,Dunne,Gama} and also \cite{Vasiliev}, and for the massless scalar model \cite{bekaert}. 
Its prominent property, as opposed to the massless one\footnote{The free massless Majorana model is plagued by 
a sign ambiguity in the definition of the partition function, \cite{Witten}. This should not be the case 
for the massive model. This problem is anyhow under investigation.}, is that
the fermion mass parameter $m$ 
breaks parity invariance, and this feature has nontrivial consequences even when
$m\to 0$.
In this paper we intend to analyze it more in depth. We will couple it to
various external sources, not only to a gauge field and a metric, but also to
higher tensor fields.  

We are interested in the one-loop effective action, in particular in the local part of its UV and
IR limits. These contributions are originated by contact terms of the correlators (for related
aspects concerning contact terms, see \cite{BL,BGL,BGLBled13,BDL}).
To do so we evaluate the 2-point correlators, and in some cases
also the 3-point correlators, of
various currents. Our method of calculation is based on Feynman diagrams and
dimensional regularization.
Eventually we take the limit of high and low energy compared to the mass $m$ of
the fermion.
In this way we recover some well-known results, \cite{Vuorio,Closset,Dunne}, and others which are perhaps not
so well-known: in the even parity sector the correlators are those (conformal
covariant) expected for the a free massless theory; in the odd parity sector the
IR limit of the 
effective action coincides with the gauge and gravity Chern-Simons (CS) action,
but also the UV limit lends itself to a similar interpretation provided we use a
suitable scaling limit. We also
couple the same theory to higher spin symmetric fields. The result we
obtain in this case for the spin 3 current in the UV limit is a
generalized CS action. We recover in this way theories proposed long ago
from a completely different point of view, \cite{pope}. In the IR limit we obtain a different
higher spin action.

We remark that in general the IR and UV correlators in the even sector are non-local, while 
the correlators in the odd-parity sector are local, i.e. made of contact terms (for related 
aspects, see \cite{Closset}).

Apart from the final results we find other interesting things in our analysis.
For instance the odd parity correlators we find as intermediate results are
conformal invariant at the fixed point. However, although we obtain 
them by taking limits of a free field theory, these correlators cannot be
obtained from any known free
field theory (using the Wick theorem). Another interesting aspect is connected
to the breaking of gauge or 
diffeomorphism symmetry in the process of taking the IR and UV limits in three-point functions. Although
we use analytic regularization, 
when taking these limits we cannot prevent a breaking of symmetry in the
correlators. They have to be `repaired' by adding suitable counterterms to the
effective action.

The paper is organized as follows. The next section is preparatory, we introduce the notation,
define the higher spin currents and the generating functions for n-point correlators.
Section 3 is devoted to two-point functions of gauge currents, of the e.m. tensor and of the spin three 
currents. In particular the local odd-parity action extracted from these 
correlators in the UV limit is identified with an action first introduced in ref. \cite{pope}. Section 4
is auxiliary: 
we discuss CS actions and their invariance analyzed with the tool of perturbative cohomology.
Section 5 is devoted to the 
three-point functions of currents, and to the rather complicated issue of conservation. In section 
6 we analyze three-point functions of the e.m. tensor and their IR and UV limits. Finally section 7 contain our
conclusions. Several Appendices are devoted to particular issues, to introduce auxiliary material 
or to show explicit calculations.

\section{The 3d massive fermion model coupled to external sources}
\label{sec:3dmassive}

The simplest model is that of a Dirac fermion\footnote{The minimal representation of the Lorentz group in 3d is 
a real Majorana fermion. A Dirac fermion is a complex combination of two Majorana fermions. The action for 
a Majorana fermion is $\frac 12$ of (\ref{actionA}).}  coupled to a gauge field. The action is 
\be \label{actionA}
S[A]=\int
d^{3}x\,\left[i\bar{\psi} \gamma^{\mu}D_{\mu}\psi-m\bar{\psi}
\psi\right],\quad D_{\mu}=
\partial_{\mu}+ A_\mu,
\ee
where $A_\mu= A_\mu^a(x) T^a$ and $T^a$ are the generators of a gauge algebra in
a given representation determined by $\psi$. We will use the antihermitean
convention, so
$[T^a,T^b]=f^{abc} T^c$, and the normalization ${\rm tr}(T^a T^b)=\delta^{ab}$. 

The current
\be
J^a_\mu(x) = \bar \psi \gamma_\mu T^a \psi\label{Jmu}
\ee
is (classically) covariantly conserved on shell as a consequence of the gauge
invariance of (\ref{actionA})
\be \label{currentconserv}
(DJ)^a = (\del^\mu \delta^{ac} + f^{abc} A^{b\mu})J_\mu^c=0.
\ee

The next example involves the coupling to gravity
\begin{equation}
S[g]=\int
d^{3}x\,e\left[i\bar{\psi}E_{a}^{\mu}\gamma^{a}\nabla_{\mu}\psi-m\bar{\psi}
\psi\right],\quad\nabla_{\mu}=
\partial_{\mu}+\frac{1}{2}\omega_{\mu
bc}\Sigma^{bc},\quad\Sigma^{bc}=\frac{1}{4}\left[\gamma^{b},\gamma^{c}\right]
.\label{actiong}
\end{equation}
The corresponding energy momentum tensor
\begin{equation} \label{EMtensor}
T_{\mu\nu}=\frac{i}{4}\bar{\psi}\left(\gamma_{\mu}\stackrel{\leftrightarrow}{
\partial}_{\nu}+\gamma_{\nu}\stackrel{\leftrightarrow}{\partial}_{\mu}
\right)\psi
\end{equation}
is covariantly conserved on shell as a consequence of the diffeomorphism
invariance of the action,
\be \label{Tconserv}
\nabla^\mu T_{\mu\nu}(x)=0.
\ee

However we can couple the fermions to more general fields. Consider the free
action
\be
S=\int
d^{3}x\,\left[i\bar{\psi} \gamma^{\mu}\del_{\mu}\psi-m\bar{\psi}
\psi\right], \label{actionfree}
\ee
and the spin three conserved current 
\begin{eqnarray} \label{J3}
J_{\mu_1\mu_2\mu_3} &=& \frac 12 \bar \psi \gamma_{(\mu_1} \partial_{\mu_2}
\partial_{\mu_3)}\psi+\frac 12  \partial_{(\mu_1}
\partial_{\mu_2}\bar \psi \gamma_{\mu_3)} \psi 
-\frac 5{3} \partial_{(\mu_1} \bar \psi \gamma_{\mu_2}
\partial_{\mu_3)}\psi\0\\
&& +\frac 13 \eta_{(\mu_1\mu_2} \partial^\sigma \bar \psi
\gamma_{\mu_3)}
\partial_\sigma \psi -\frac {m^2}3 \eta_{(\mu_1\mu_2} \bar \psi \gamma_{\mu_3)} \psi.
\end{eqnarray}
Using the equation of motion one can prove that 
\begin{eqnarray}
&&\partial^\mu J_{\mu\nu\lambda} =0, \label{conservJ3}\\
&& J_{\mu}{}^\mu{}_\lambda = \frac 49 m \left( -i \partial_\lambda \bar \psi
\psi +i \bar \psi \partial_\lambda \psi +2 \bar\psi \gamma_\lambda \psi
\right).\label{trJ3}
\end{eqnarray}
Therefore, the spin three current (\ref{J3}) is conserved on shell and its
tracelessness is softly broken by the mass term. Similarly to the gauge field
and the metric, we can couple
the fermion $\psi$ to a new external source $b_{\mu\nu\lambda}$ by adding to
(\ref{actionfree}) the term
\begin{equation} \label{3rdordercoupling}
\int d^3x J_{\mu\nu\lambda} b^{\mu\nu\lambda}.
\end{equation}
Notice that this requires $b$ to have canonical dimension -1.
Due to the (on shell) current conservation this coupling is invariant under the
(infinitesimal) gauge transformations
\begin{equation} \label{Bgaugetransf}
\delta b_{\mu\nu\lambda} = \partial_{(\mu}\Lambda_{\nu\lambda)},
\end{equation}
where round brackets stand for symmetrization. In the limit $m\to 0$ we have
also invariance under the local transformations
\begin{equation} \label{Bweyltransf}
\delta  b_{\mu\nu\lambda}=\Lambda_{(\mu} \eta_{\nu\lambda)}, 
\end{equation}
which are usually referred to as (generalized) Weyl transformations and which 
induce the tracelessness of $J_{\mu\nu\lambda}$ in any couple of indices. 

The construction of conserved currents can be generalized as follows, see
\cite{giombi,GMPTWY}.
There is a generating function for $J^{(n)}$. Introduce the following symbols
\begin{equation}
u_\mu= \stackrel{\rightarrow}{\partial}_\mu,\quad\quad v_{\mu}= \stackrel
{\leftarrow}{\partial_\mu},\quad\quad
\langle uv\rangle= u^\mu v_\mu, \quad\quad \langle uz\rangle= u^\mu z_\mu,
\quad\quad \langle \gamma z\rangle= \gamma^\mu z_\mu, \quad\quad {\rm etc,}\0
\end{equation}
where $z^\mu$ are external parameters. Now define
\begin{equation} \label{Jxz}
J(x;z) = \sum_n J^{(n)}_{\mu_1\ldots \mu_n}z^{\mu_1}\ldots z^{\mu_n}= \bar \psi
\langle \gamma z\rangle F(u,v,z)\psi,
\end{equation}
where 
\begin{equation} \label{Fuvz}
F(u,v,z)= e^{(\langle uz\rangle-\langle vz\rangle)}\,\,f(X),\quad\quad
f(X)=\frac {\sinh {\sqrt X}}{\sqrt X}, \quad\quad X= 2\langle uv\rangle \langle
zz\rangle- 4 \langle uz\rangle\langle vz\rangle.
\end{equation}
Defining next the operator ${\cal D}= \langle (u+v)\frac {\partial}{\partial
z}\rangle$, it is easy to prove
that, using the free equation of motion,
\begin{equation} \label{conservation}
{\cal D} J(x;z)=0.
\end{equation}
Therefore all the homogeneous terms in $z$ in $J(x;z)$ are conserved if $m=0$.
If $m\neq 0$ one has to replace
$X$ with $Y=X-2m^2\langle zz\rangle$. Then we define
\begin{equation}
J_m(x;z) = \sum_n J^{(n)}_{\mu_1\ldots \mu_n}z^{\mu_1}\ldots z^{\mu_n}= \bar
\psi \langle \gamma z\rangle  e^{(\langle uz\rangle-\langle vz\rangle)}\,\,f(Y) 
\psi\label{Jmxz}
\end{equation}
and one can prove that
\begin{equation} \label{conservationm}
{\cal D} J_m(x;z)=0,
\end{equation}
with $m\neq 0$. The case $J^{(3)}$ in (\ref{Jmxz}) coincides with the third
order current introduced before. 
 
For any conserved current $J^{(n)}_{\mu_1\ldots \mu_n}$ we can introduce an
associated source field $b^{\mu_1\ldots\mu_n}$ similar to the rank three one
introduced above, with a transformation law that generalizes
(\ref{Bgaugetransf}).
However, in this regard, a remark is in order. In fact, (\ref{Bgaugetransf}) has to be understood 
as the transformation of the fluctuating field $b_{\mu\nu\lambda}$, which is the
lowest order term in the expansion 
of a field $B_{\mu\nu\lambda}= b_{\mu\nu\lambda}+\ldots$ whose background value
is 0. $b_{\mu\nu\lambda}$ plays a role similar to $h_{\mu\nu}$ in the expansion
of the metric $g_{\mu\nu}=\eta_{\mu\nu}+ h_{\mu\nu}+\ldots$ (see also Appendix
B). In order to implement full invariance we should introduce in the free action
the analog of the spin connection for $B_{\mu\nu\lambda}$ and a full covariant
conservation law would require introducing in (\ref{conservJ3}) the analog of
the Christoffel symbols.

\subsection{Generating function for effective actions}
\label{ssec:genfunct}

The generating function of the effective action of (\ref{actionA}) is
\be \label{WA}
W[A]&=&\sum_{n=1}^\infty \frac {i^{n+1}}{n!} \int \prod_{i=1}^n d^3x_i A^{a_1
\mu_1}(x_1)  \ldots A^{a_n \mu_n}(x_n) 
\langle 0|{\cal T} J_{\mu_1}^{a_1}(x_1)\ldots
J_{\mu_n}^{a_n}(x_n)|0\rangle,
\ee
where the time ordered correlators are understood to be those obtained with the
Feynman rules.
The full one-loop 1-pt correlator for $J_\mu^a$ is
\be 
\langle\!\langle J_\mu^a(x)\rangle\!\rangle &=& \frac {\delta W[A]}{\delta
A^{a\mu}(x)}\label{Jamux}\\
&=& - \sum_{n=1}^\infty \frac {i^n}{n!} \int \prod_{i=1}^n d^3x_i  A^{a_1
\mu_1}(x_1)  \ldots A^{a_n \mu_n}(x_n) \langle 0|{\cal T}J_\mu^a(x)
J_{\mu_1}^{a_1}(x_1)
\ldots J_{\mu_n}^{a_n}(x_n)|0\rangle . \0
\ee
Later on we will need also the one-loop conservation 
\be \label{qconsrv}
(D^\mu \langle\!\langle J_\mu(x)\rangle\!\rangle)^a= \del^\mu \langle\!\langle
J_\mu^a(x)\rangle\!\rangle + f^{abc} A_\mu^b(x) \langle\!\langle J^{\mu
c}(x)\rangle\!\rangle=0 .
\ee
We can easily generalize this to the case of higher tensor currents $J^{(p)}$.
The generating function is
\begin{multline} \label{Wap}
W^{(p)}[a] =\sum_{n=1}^\infty \frac {i^{n+1}}{n!} \int \prod_{i=1}^n d^3x_i
a^{\mu_{11}\ldots \mu_{1p}}(x_1)  \ldots a^{\mu_{n1}\ldots \mu_{np}}(x_n) \\ 
\times \langle 0|{\cal T} J^{(p)}_{\mu_{11}\ldots\mu_{1p}}(x_1)\ldots
J^{(p)}_{\mu_{n1}\ldots\mu_{np}}(x_n)|0\rangle . 
\end{multline}
In particular $a_{\mu\nu}= h_{\mu\nu}$ and $J^{(2)}_{\mu\nu}=T_{\mu\nu}$,  and
$a_{\mu\nu\lambda}= b_{\mu\nu\lambda}$.
The full one-loop 1-pt correlator for $J_\mu^a$ is
\begin{multline} \label{Japmux}
\langle\!\langle J^{(p)}_{\mu_1\ldots \mu_p}(x)\rangle\!\rangle = \frac{\delta W[a,p]}{\delta a^{\mu_1\ldots \mu_p}(x)}
= - \sum_{n=1}^\infty \frac {i^n}{n!} \int \prod_{i=1}^n d^3x_i 
a^{\mu_{11}\ldots \mu_{1p}}(x_1)  \ldots a^{\mu_{n1}\ldots \mu_{np}}(x_n) \\
\times \langle 0|{\cal T}  J^{(p)}_{\mu_{1}\ldots\mu_{p}}(x)J^{(p)}_{\mu_{11}\ldots\mu_{1p}}(x_1)\ldots
J^{(p)}_{\mu_{n1}\ldots\mu_{np}}(x_n)|0\rangle . 
\end{multline}
The full one-loop conservation law for the energy-momentum tensor is
\be \label{emconserv}
\nabla^\mu\langle\!\langle T_{\mu\nu}(x)\rangle\!\rangle=0.
\ee
A similar covariant conservation should be written also for the other currents,
but in this paper for $p>2$ we will content ourselves with the lowest nontrivial 
order in which the conservation law reduces to
\be \label{Jpconserv}
\del^{\mu_1} \langle\!\langle J^{(p)}_{\mu_1\ldots
\mu_p}(x)\rangle\!\rangle=0.
\ee

\vskip 0.3cm

{\bf Warning}. One must be careful when applying the previous formulas for generating functions.
If the expression $\langle 0|{\cal T} J^{(p)}_{\mu_{11}\ldots\mu_{1p}}(x_1)\ldots
J^{(p)}_{\mu_{n1}\ldots\mu_{np}}(x_n)|0\rangle$ in (\ref{Wap}) is meant to denote
the $n$-th point-function calculated by using Feynman diagrams, a factor $i^n$ is already
included in the diagram themselves and so it should be dropped in (\ref{Wap}). When the
current is the energy-momentum tensor an additional precaution is necessary: the factor
$\frac {i^{n+1}}{n!}$ must be replaced by $\frac {i}{2^n n!}$.    
The factor 
$\frac 1{2^n}$  is motivated by the fact that when we expand the action 
$$S[\eta+h]= S[\eta] + \int d^dx \frac {\delta S}{\delta g^{\mu\nu}}\Big{\vert}_{g=\eta} h^{\mu\nu}+\cdots,$$ 
the factor $ \frac {\delta S}{\delta g^{\mu\nu}}\Big{\vert}_{g=\eta}= \frac 12 T_{\mu\nu}$. 
Another consequence of this fact will be that the presence of vertices with one graviton in 
Feynman diagrams will correspond to insertions of the operator $\frac{1}{2}T_{\mu\nu}$ 
in correlation functions.

\subsection{General structure of 2-point functions of currents}
\label{ssec:gen2pt}

In order to compute the generating function (effective action) $W$ we will
proceed in the next section to
evaluate 2-point and 3-point correlators using the Feynman diagram approach. It is
however possible to derive 
their general structure on the basis of covariance. In this subsection we will
analyze the general form
of 2-point correlators. 

As long as 2-point correlators of currents are involved the conservation law is
simply
represented by the vanishing of the correlator divergence:
\be \label{2ptconserv}
\partial^{\mu_1} \langle0| {\cal T} J^{(p)}_{\mu_1\ldots \mu_p}(x) J^{(p)}_{\nu_1\ldots \nu_p}(y)|0\rangle=0.
\ee
Using Poincar\'e covariance and this equation we can obtain the general form of
the correlators in momentum space in terms of distinct tensorial structures and
form factors. Denoting by
\be \label{JJtilde}
\tilde J_{\mu_1\ldots \mu_p,\nu_1\ldots \nu_p}(k)=\langle \tilde
J^{(p)}_{\mu_1\ldots \mu_p}(k) \tilde J^{(p)}_{\nu_1\ldots
\nu_p}(-k)\rangle
\ee
the Fourier transform of the 2-point function, the conservation is simply
represented by the contraction of $\tilde F_{\mu\ldots}$ with $k^\mu$:
\be \label{2ptconservk}
k^{\mu_1} \tilde J_{\mu_1\ldots \mu_p,\nu_1\ldots \nu_p}(k)=0.
\ee
The result is as follows. For 1-currents we have
\be
\tilde J^{ab}_{\mu\nu}(k) = \langle \tilde J^{a}_{\mu}(k) \tilde
J^{b}_{\nu}(-k)\rangle=\delta^{ab}\left[
\tau\left(\frac {k^2}{m^2}\right)\frac{k_{\mu}k_{\nu}-k^{2}\eta_{\mu\nu}}{16
|k|}+\kappa\left( \frac{k^2}{m^2} \right)\frac{k^{\tau}\epsilon_{\tau\mu\nu}}{2\pi}\right].\label{JaJb}
\ee
where $|k|=\sqrt{k^2}$ and $\tau,\kappa$ are model dependent form factors.

The most general 2-point function for the energy-momentum tensor has the form
 \begin{eqnarray}
\left\langle \tilde T_{\mu\nu}\left(k\right)\tilde
T_{\rho\sigma}\left(-k\right)\right\rangle 
& = &
\frac{\tau_{g}\left( k^2/m^2 \right)}{|k|}\left(k_{\mu}k_{\nu}
-\eta_{\mu\nu}k^{2}\right)\left(k_{\rho}k_{\sigma}-\eta_{\rho\sigma}k^{2}
\right)\nonumber \\
 &  &
+\frac{\tau'_{g}\left( k^2/m^2 \right)}{|k|}\left[\left(k_{\mu}k_{
\rho}-\eta_{\mu\rho}k^{2}\right)\left(k_{\nu}k_{\sigma}-\eta_{\nu\sigma}k^{2}
\right)+\mu\leftrightarrow\nu\right]\label{eq:Conserved2ptEM}\\
&  &
+\frac{\kappa_{g}\left( k^2/m^2 \right)}{192\pi}
\left[
\left(\epsilon_{\mu\rho\tau} k^{\tau}\left(k_{\nu}k_{\sigma}-\eta_{\nu\sigma}k^{2}\right)+ \rho \leftrightarrow \sigma \right) +\mu\leftrightarrow\nu
\right].\nonumber 
\end{eqnarray}
where $\tau_g, \tau'_g$ and $\kappa_g$ are model-dependent form-factors. 
Vanishing of traces over $(\mu\nu)$ or $(\rho\sigma)$ requires $\tau_g + \tau'_g = 0$. 
Both here and in the previous case, the notation, the signs and the numerical factors 
are made to match our definition with the ones used in \cite{Closset}.
\footnote{Except that we work in spacetime with Lorentzian signature $(+--)$.}

As for the order 3 tensor currents the most general form of the 2-point function in momentum representation is
\begin{multline}
\langle \tilde J_{\mu_1\mu_2\mu_3}(k)  \tilde J_{\nu_1\nu_2\nu_3}(-k)\rangle = 
\tau_b \left(\frac{k^2}{m^2}\right) |k|^5 \pi_{\mu_1 \mu_2}  \pi_{\mu_3 \nu_1} \pi_{\nu_2 \nu_3} 
+ \tau'_b \left(\frac{k^2}{m^2}\right) |k|^5 \pi_{\mu_1 \nu_1}  \pi_{\mu_2 \nu_2} \pi_{\mu_3 \nu_3} \\
+ k^4 \epsilon_{\mu_1 \nu_1 \sigma} k^\sigma\left[\kappa_b \left(\frac{k^2}{m^2}\right)  \pi_{\mu_2 \mu_3} \pi_{\nu_2 \nu_3} 
+ \kappa'_b \left(\frac{k^2}{m^2}\right)  \pi_{\mu_2 \nu_2} \pi_{\mu_3 \nu_3}\right],
\end{multline}
where complete symmetrisation of the indices $(\mu_1,\mu_2,\mu_3)$ and $(\nu_1,\nu_2,\nu_3)$ 
is implicit\footnote{When we say that the complete symmetrisation is implicit it means 
that one should understand, for instance
\be
\pi_{\mu_1 \mu_2}  \pi_{\mu_3 \nu_1} \pi_{\nu_2 \nu_3}  \rightarrow \frac{1}{9} 
\left[ \pi_{\mu_1 \mu_2}  \pi_{\mu_3 \nu_1} \pi_{\nu_2 \nu_3} + \pi_{\mu_1 \mu_3}  
\pi_{\mu_2 \nu_1} \pi_{\nu_2 \nu_3}  + \dots \right]. \0
\ee
} and
\begin{equation}
\pi_{\mu \nu} = \eta_{\mu\nu} - \frac{k_{\mu}k_{\nu}}{k^2}
\end{equation}
is the transverse projector. This expression is, by construction, conserved but not traceless. Vanishing of traces requires 
\begin{equation}
4\tau_b +3\tau'_b =0,\quad\quad 4\kappa_b +\kappa'_b =0. \label{trace0cond}
\end{equation}
 
\section{Two-point functions}
\label{sec:2ptfunct}

In this section we compute the the 2-point function of spin 1, 2 and 3 currents
using Feynman diagrams with finite mass $m$. Then we take the limit $m\to 0$ or
$m\to \infty$ with respect to the total energy of the process, i.e. the UV and
IR limit of the 2-point functions, respectively. These are expected to correspond to
2-point functions of conformal field theories at the relevant fixed points. We will
be mostly
interested in the odd parity part of the correlators, because in the UV and IR
limit they give rise to
local effective actions, but occasionally we will also consider the even parity
part.

\subsection{Two-point function of the current $J_\mu^a(x)$}
\label{ssec:J1J1}

This case has been treated in \cite{BL}, therefore we will be brief. The only
contribution comes from the 
bubble diagram with external momentum $k$ and momentum $p$ in the fermion loop.
In momentum representation we have
\begin{multline}\label{Jmunuab}
\tilde J_{\mu\nu}^{ab}(k) = - \int \frac{d^3p}{(2\pi)^3} {\rm Tr}
\left(\gamma_\mu T^a \frac 1{\slashed{p} -m }\gamma_\nu T^b  
\frac 1{\slashed{p}-\slashed{k} -m }\right) = -2 \delta^{ab} \\
\times \int\frac{d^3p}{(2\pi)^3}\frac {p_\nu(p-k)_\mu
-p\!\cdot\!(p-k)\eta_{\mu\nu} + p_\mu(p-k)_\nu+im \epsilon_{\mu\nu\sigma}
k^\sigma +m^2 \eta_{\mu\nu}}{(p^2-m^2)((p-k)^2-m^2)}
\end{multline}
For the even parity part we get
\be\label{Jmunueven}
\tilde J_{\mu\nu }^{ab(even)}(k) = \frac{2 i}{\pi} \delta^{ab} \left[ \left( 1 
+ \frac{4 m^2}{k^2} \right) \arctanh\left( \frac{|k|}{2 |m|} \right) 
- \frac{2 |m|}{|k|} \right]
\frac{k_\mu k_\nu - k^2 \eta_{\mu \nu}}{16|k|},
\ee
while for the odd parity part we get
\be\label{Jmunuodd}
\tilde J_{\mu\nu }^{ab(odd)}(k)  =  \
\frac {1}{2\pi} \delta^{ab} \epsilon_{\mu\nu\sigma} k^\sigma \, \frac{m}{|k|} \arctanh\left(\frac{|k|}{2 |m|}\right)
\ee
where $|k|= \sqrt{k^2}$.  The conservation law (\ref{2ptconservk}) is readily seen to be satisfied.
In the following we are going to consider the IR and UV limit of the expressions (\ref{Jmunueven}) 
and (\ref{Jmunuodd}) and it is important to remark that we have two possibilities here: we may consider a 
timelike momentum ($k^2 > 0$) or a spacelike one ($k^2 < 0$). In the first case, we must notice that 
the function $\ArcTanh{\frac{|k|}{2 |m|}}$ has branch-cuts on the real axis for $\frac{|k|}{2 |m|} > 1$ 
and it acquires an imaginary part. On the other hand, if we consider spacelike momenta, we will have 
$\ArcTanh{\frac{i |k|}{2 |m|}} = i \arctan \left( \frac{|k|}{2 |m|} \right)$ and $\arctan \left( \frac{|k|}{2 |m|} \right)$ 
is real on the real axis. The region of spacelike momenta reproduces the Euclidean correlators. Throughout 
this paper we will always consider UV and IR limit as being respectively the limits of very large or very 
small spacelike momentum with respect to the mass scale $m$.
In these two limits we get
\be \label{JmunuEvenIRUV}
\tilde J_{\mu\nu }^{ab(even)}(k)= \frac {i}{8\pi} \delta^{ab}\frac{k_\mu k_\nu - k^2 \eta_{\mu \nu} }{|k|}
\left\{
\begin{matrix} 
	\frac{2 |k|}{3 |m|}	&	\quad {\rm IR}\\                                                                
	\frac{\pi}{2}	&	\quad {\rm UV}
\end{matrix}
\right. ,
\ee
\be \label{JmunuIRUV}
\tilde J_{\mu\nu }^{ab(odd)}(k)= \frac 1{2\pi} \delta^{ab}
\epsilon_{\mu\nu\sigma} k^\sigma 
\left\{
\begin{matrix} 
	\frac 12 \frac {m}{|m|}	&	\quad {\rm IR}\\                                                                
        \frac {\pi}{2} \frac{m}{|k|}	&	\quad {\rm UV}                                                                             
 \end{matrix}
 \right. .
\ee
The UV limit is actually vanishing in the odd case (this is also the case for all the 2-point
functions we will meet in the following). However we can consider a
model made of $N$ identical copies of free fermions coupled to the same gauge
field. Then the result (\ref{JmunuIRUV}) would be
\be \label{JmunuUV}
\tilde J_{\mu\nu }^{ab(odd)}(k)= \frac { N}4 \delta^{ab}
\epsilon_{\mu\nu\sigma} k^\sigma  \frac{m}{|k|}.
\ee
In this case we can consider the scaling limit $ \frac{m}{|k|} \to 0$ and $N\to
\infty$ in such a way that $N  \frac{m}{|k|}$ is fixed. Then the UV limit
(\ref{JmunuUV}) becomes nonvanishing.

Fourier transforming (\ref{JmunuIRUV}) and inserting the result in the
generating function (\ref{WA})
we get the first (lowest order) term of the CS action  
\be
CS &=& \frac {\kappa}{4\pi} \int d^3x {\rm Tr} \left( A\wedge dA + \frac 23
A\wedge A \wedge A\right)\label{gaugeCS}\\
&=& \frac { \kappa}{4\pi} \int d^3x \epsilon^{\mu\nu\lambda} \left(
A^a_\mu \partial_\nu A^a_\lambda 
+\frac{1}{3} f^{abc} A_\mu^a A_\nu^b A_\lambda^c\right).\0
\ee
In particular, from (\ref{JmunuIRUV}) we see that in the IR limit $\kappa=\pm \frac{1}{2}$. 
The CS action (\ref{gaugeCS}) is invariant not only under the infinitesimal gauge transformations
\be
\delta A = d \lambda + [A,\lambda], \quad\quad \lambda = \lambda^a(x) T^a,
\label{Agaugetransf}
\ee
but also under large gauge transformations when $\kappa \in \mathbb{Z}$. From (\ref{JmunuIRUV}) follows 
that $\kappa_{UV} = 0$ and $\kappa_{IR} = \pm 1/2$, which suggests that the gauge symmetry 
is broken unless there is an even number of fermions. A further discussion of this phenomenon can be found in \cite{Closset}.

\subsection{Two-point function of the e.m. tensor}
\label{ssec:TT}

The lowest term of the effective action in an expansion in $h_{\mu\nu}$ come
from the two-point function of the e.m. tensor. So we now set out to compute the
latter. The correlators of the e.m. tensor will be denoted with the letter $\tilde T$ instead of $\tilde J$.
The Feynman propagator and vertices are given in Appendix \ref{sec:invariances}. For simplicity from now on
we assume $m>0$.

The bubble diagram (one graviton entering and one graviton exiting with momentum
$k$, one fermionic loop) contribution to the e.m. two-point function is given in
momentum space by
\be\label{Tmnlr}
\tilde T_{\mu\nu\lambda\rho}(k) = - \frac 1{64} \int \!\! \frac {d^3p}{(2\pi)^3}
{\rm Tr}\left(\frac 1{\slashed{p}-m}(2p-k)_\mu \gamma_\nu \frac
1{\slashed{p}-\slashed{k}-m}(2p-k)_\lambda \gamma_\rho\right),
\ee
where symmetrization over $\left( \mu, \nu \right)$ and $\left( \lambda, \rho \right)$ will be always implicit.
\subsubsection{The odd parity part}

The odd-parity part of (\ref{Tmnlr}) is
\begin{equation}\label{Tmnlrodd}
\tilde T^{(odd)}_{\mu\nu\lambda\rho}(k) = \frac {im}{32} \! \int_0^1 \!\! dx \! \int \!\! \frac{d^3p}{(2\pi)^3}
\epsilon_{\sigma \nu\rho}\, k^\sigma  \, \frac
{(2p+(2x-1)k)_\mu \,(2p+(2x-1)k)_\lambda}{[p^2-m^2 +x(1-x)k^2]^2}.
\end{equation}
The evaluation of this integral is described in detail in Appendix
\ref{sec:usefulint}.
The result is
\begin{multline} \label{Tmnlrfinal}
\tilde T^{(odd)}_{\mu\nu\lambda\rho}(k) =  - \frac{3 m}{4 |k|} \left[ \left( 1 - \frac{4 m^2}{k^2} \right) 
\ArcTanh{\frac{|k|}{2|m|}} + \frac{2 |m|}{|k|} \right]  \frac{\epsilon_{\mu \lambda \sigma} 
k^\sigma \left( k_\nu k_\rho - \eta_{\nu \rho} k^2 \right)}{192\pi} - \\
-\frac{\mathrm{sign}(m) |m|^2}{64\pi}\epsilon_{\mu \lambda \sigma} k^\sigma \eta_{\nu \rho} .
\end{multline}
A surprising feature of (\ref{Tmnlrfinal}) is that if we contract it with
$k^\mu$ we do not get zero. Let us look closer into this problem.

\subsubsection{The divergence of the e.m. tensor: odd-parity part}

To see whether the expression of the one-loop effective action is the legitimate
one, one must verify that the procedure to obtain it does not break diffeomorphism
invariance. The bubble diagram  contribution to the divergence of the e.m.
tensor is 
\begin{eqnarray} \label{kTnlr}
k^\mu \tilde T_{\mu\nu\lambda\rho}(k) &=& - \frac 1{64} \int\frac
{d^3p}{(2\pi)^3}\left[ {\rm Tr}\left(\frac 1{\slashed{p}-m}(2p-k)\cdot k
\,\gamma_\nu \frac 1{\slashed{p}-\slashed{k}-m}(2p-k)_\lambda
\gamma_\rho\right)\right.\nonumber\\
&& + \left. {\rm Tr}\left(\frac 1{\slashed{p}-m}(2p-k)_\nu\, \slashed{k} \frac
1{\slashed{p}-\slashed{k}-m}(2p-k)_\lambda \gamma_\rho\right)
\right]+\left(\lambda\leftrightarrow \rho\right).
\end{eqnarray}
Repeating the same calculation as above one finally finds
\begin{eqnarray} \label{diffanommodd}
k^\mu\tilde T^{(odd)}_{\mu\nu\lambda\rho}(k) 
&=& - \frac {\mathrm{sign}(m) |m|^2}{64\pi}\epsilon_{\sigma \nu\rho}\, k^\sigma k_\lambda + 
\left(\lambda\leftrightarrow \rho\right).
\end{eqnarray}
This is a local expression. It corresponds to the anomaly
\begin{equation} \label{Deltaxi}
\Delta_\xi = - \frac {\mathrm{sign}(m) |m|^2}{32\pi}\int \epsilon_{\sigma \nu\rho} \xi^\nu
\partial^\sigma \partial_\lambda h^{\lambda\rho}.
\end{equation}
The counterterm to cancel it is
\begin{equation} \label{Cxi}
{\cal C}= \frac {\mathrm{sign}(m) |m|^2}{64\pi}\int \epsilon_{\sigma \nu\rho} h_{\lambda}^\nu \partial^\sigma    h^{\lambda\rho}.
\end{equation}
Once this is done the final result is 
\begin{eqnarray}
\langle T_{\mu\nu}(k)\, T_{\lambda\rho}(-k)\rangle_{odd} &=& 
\frac{\kappa_g(k^2/m^2)}{192\pi}\, \epsilon_{\sigma \nu\rho}\, k^\sigma
\left( k_\mu k_\lambda - k^2 \eta_{\mu\lambda} \right) +
\left(\begin{matrix}{\mu\leftrightarrow \nu}\\{\lambda\leftrightarrow
\rho}\end{matrix}\right) \qquad \label{twopointoddL}
\end{eqnarray}
with
\begin{equation} \label{kappag}
\kappa_g(k^2/m^2) =  - \frac{3 m}{|k|} \left[ \left( 1 - \frac{4 m^2}{k^2} \right) \ArcTanh{\frac{|k|}{2 |m|}} + \frac{2 |m|}{|k|} \right].
\end{equation}
Now (\ref{twopointoddL}) is conserved and traceless. To obtain (\ref{kappag}) we have to recall that
\be
 \tilde T_{\mu\nu\lambda\rho}(k) = \frac{1}{4}\langle T_{\mu\nu}(k)\, T_{\lambda\rho}(-k)\rangle,
\ee
which was explained in the warning of section \ref{ssec:genfunct}.
To complete the discussion we should also take into account a tadpole graph which might
contribute to the two-point function. With the vertex $V_{ggff}$ it is in fact
possible to construct such a graph. It yields the contribution
\be 
\frac{3}{32\pi} \mathrm{sign}(m)|m|^2 \, t_{\mu\nu\lambda\rho\sigma} \, k^\sigma .
\ee
This term violates conservation, just as the previous (\ref{diffanommodd}), but it has a different
coefficient. So it must be subtracted in the same way.

\subsubsection{The UV and IR limit}

Let us set  $\lim_{\left|\frac {m}{k}\right| \to 0} \kappa_g= \kappa_{UV}$, and
$\lim_{\left|\frac {k}{m}\right| \to 0} \kappa_g= \kappa_{IR}$. We get
\be \label{IRUVlimit}
\kappa_{IR}=\frac m{|m|},\quad\quad  \kappa_{UV}= \frac 32 \pi \frac{m}{|k|}=0 + \mathcal{O}\left( \left|\frac {m}{k}\right| \right).
\ee
As before for the gauge case, in the UV limit we can get a finite result by
considering a system of $N$ identical fermions. Then the above 2-point function
gets 
multiplied by $N$. In the UV limit, $\left|\frac {m}{k}\right|\to 0$,  we can
consider the scaling limit $N\to \infty, \left|\frac {m}{k}\right|\to 0$ such that 
\begin{equation}
\lambda= N  \frac m{|k|}\label{lambda}
\end{equation}
is fixed and finite. In this limit 
\begin{equation}
 \lim_{N\to\infty,\left|\frac {m}{k}\right|\to 0 } N \kappa_g(k)=  \frac{3\pi}{2} \frac {m}{|m|}  \lambda.
\label{scalinglimit}
\end{equation}
For a unified treatment let us call both UV and IR limits of $\kappa_g$ simply
$\kappa$. 
In such limits contribution to the parity odd part of the effective action can
be easily reconstructed
by 
\begin{equation}
S_{\mathrm{eff}}^{(odd)} = \frac{\kappa}{192\pi} \int d^3x\,
\epsilon_{\sigma\nu\rho}\,
 h^{\mu\nu}\, \partial^\sigma ( \partial_\mu \partial_\lambda -
\eta_{\mu\lambda} \Box)\, h^{\lambda\rho}.
\end{equation}
This exactly corresponds to a gravitational CS term in 3d, for which at the
quadratic order in $h_{\mu\nu}$ we have
\begin{eqnarray}
CS&=&-\frac{\kappa}{96\pi}\int d^3x\, \epsilon^{\mu\nu\lambda}
\left(\partial_\mu\omega_\nu^{ab}
\omega_{\lambda b a}+ \frac 23 \omega_{\mu a}{}^b \omega_{\nu b}{}^c
\omega_{\lambda c}{}^a\right) \label{gravityCS}\\
&=& \frac{\kappa}{192\pi} \int d^3x\,  \epsilon_{\sigma\nu\rho}\, h^{\lambda
\rho}\left( \partial^\sigma
\partial_\lambda \partial_b h^{b\nu}-\partial^\sigma \square h_\lambda^\nu
\right)+ \ldots\nonumber
\end{eqnarray}
Once again we note that the topological arguments combined with path
integral quantization force $\kappa$ to be an integer 
($\kappa \in \mathbb{Z}$). From $\kappa_{IR}=\pm 1$ we see that the quantum
contribution
to the parity-odd part of the effective action in the
IR is given by the local gravitational CS term, with the minimal (positive or
negative unit) coupling
constant. The constant $\frac{3\pi}{2}\lambda$ in the UV has of course to be integer in order for the 
action to be well defined also for large gauge transformations. Finally we recall that the CS Lagrangian 
is diffeomorphism and  Weyl invariant up to a total derivative. However, note that for the Majorana 
fermion one would obtain half of the result as for the Dirac fermion, i.e. $\kappa_{IR}=\pm 1/2$.

\subsubsection{Two-point function: even parity part}
 
Although in this paper we are mostly interested in the odd-parity amplitudes,
for completeness
in the following we calculate also the even parity part of the e.m. tensor 2-point
correlator.

The even parity part of the two-point function comes from the bubble
diagram alone, eq.(\ref{Tmnlr}).
Proceeding in the same way as above one gets 
\be
\begin{split}\label{TmnlrL}
\tilde T^{(even)}_{\mu\nu\lambda\rho}(k) & =
 \frac{1}{4} \tau_{g}\left( \frac {k^2}{m^2} \right)\frac{1}{|k|}\left(k_{\mu}k_{\nu}-\eta_{\mu\nu}k^{2}\right)
\left(k_{\lambda}k_{\rho}-\eta_{\lambda\rho}k^{2}\right)	\\
 &  + \frac{1}{4} \tau'_{g}\left( \frac {k^2}{m^2} \right) \frac{1}{ |k|}
 \left[\left(k_{\mu}k_{\lambda}-\eta_{\mu\lambda}k^{2}\right)\left(k_{\nu}k_{\rho}-\eta_{\nu\rho}k^{2}\right)+\mu\leftrightarrow\nu \right]	\\
& -\frac{i m^3}{48 \pi }\left( \eta_{\mu\lambda} \eta_{\nu\rho}+ \eta_{\mu\rho} \eta_{\nu\lambda} + 2 \eta_{\mu\nu} \eta_{\lambda\rho}\right),  
\end{split}
\ee
where
\be \label{tau_g}
\tau_{g}\left( \frac {k^2}{m^2} \right) = \frac{i}{64 \pi |k|^3}\left[ |m| k^2 + 4 |m|^3 - 
\frac{\left( k^2 - 4m^2 \right)^2}{2|k|} \ArcTanh{\frac{|k|}{2|m|}} \right],
\ee
\be \label{tau_g_prime}
\tau'_{g}\left( \frac {k^2}{m^2} \right) = \frac{i}{64 \pi |k|^3}\left[ 4 |m|^3 - |m| k^2 + 
\frac{\left( k^4 - 16 m^4 \right)}{2|k|} \ArcTanh{\frac{|k|}{2|m|}} \right].
\ee
Saturating (\ref{TmnlrL}) with $k^\mu$ we find
\begin{equation} \label{kmTmnlr}
k^\mu  \tilde T^{(even)}_{\mu\nu\lambda\rho}(k)= - \frac {i m^3}{48
\pi}\left(k_\lambda \eta_{\nu\rho} +k_\rho \eta_{\nu\lambda} 
+2k_\nu\eta_{\lambda\rho}\right).
\end{equation}
The same result can be obtained directly from the even part of (\ref{kTnlr}).
The term (\ref{kmTmnlr}) is local and corresponds to an anomaly proportional to
\begin{equation} \label{A}
{\cal A}_{\xi} = \int \xi^\nu \left(\partial_\lambda h^\lambda_\nu+ \partial_\nu
h\right) .
\end{equation}
This can be eliminated by subtracting the counterterm
\begin{equation} \label{C} 
{\cal C} = -\frac 12 \int (h^{\lambda\nu} h_{\lambda\nu}+h^2).
\end{equation}
After this we can write
\be
\begin{split} \label{TTAfterCT}
\langle T_{\mu\nu}\left(k\right) T_{\lambda\rho}\left(-k\right)\rangle_{even} & = \tau_{g}\left( \frac {k^2}{m^2} \right)
\frac{1}{|k|}\left(k_{\mu}k_{\nu}-\eta_{\mu\nu}k^{2}\right)
\left(k_{\lambda}k_{\rho}-\eta_{\lambda\rho}k^{2}\right)	\\
 &  + \tau'_{g}\left( \frac {k^2}{m^2} \right) \frac{1}{ |k|} \left[\left(k_{\mu}k_{\lambda}-\eta_{\mu\lambda}k^{2}\right)
\left(k_{\nu}k_{\rho}-\eta_{\nu\rho}k^{2}\right)+\mu\leftrightarrow\nu \right].
\end{split}
\ee
The UV limit gives
\begin{equation} \label{tauUVlimit}
 \lim_{\left| \frac{k}{m} \right|\to \infty} \tau_g= -\lim_{\left| \frac{k}{m} \right|\to \infty} \tau'_g =\frac{1}{256},
\end{equation}
so that in this limit
\begin{eqnarray} \label{TTevenUV} 
\langle T_{\mu\nu}\left(k\right)
T_{\lambda\rho}\left(-k\right)\rangle^{UV} _{even}
& = &
-\frac {i}{256}\frac 1{|k|}\Big{(}\left(k_{\mu}k_{\nu}
-\eta_{\mu\nu}k^{2}\right)\left(k_{\rho}k_{\lambda}-\eta_{\rho\lambda}k^{2}
\right) \nonumber \\
 &  &
-\left[\left(k_{\mu}k_{\rho}-\eta_{\mu\rho}k^{2}\right)\left(k_{\nu}k_{\lambda}
-\eta_{\nu\lambda}k^{2}
\right)+\mu\leftrightarrow\nu\right]\Big{)}.
\end{eqnarray}
This represents the two-point function of a CFT in 3d, which is a free theory, the massless limit of the massive fermion theory we are studying.

The IR limit of the form factors (\ref{tau_g}) and (\ref{tau_g_prime}) is
\begin{eqnarray} \label{IRTTformfactors}
\tau_g & = & \frac{1}{24\pi}  \left|\frac{m}{k}\right| + \mathcal{O}\left( \left|\frac{k}{m}\right| \right),\\
\tau'_g & = & - \frac{1}{48\pi}  \left|\frac{m}{k}\right| + \mathcal{O}\left( \left|\frac{k}{m}\right| \right).
\end{eqnarray}
In this limit we have
\begin{multline} \label{IRlimitTT}
\langle T_{\mu\nu}(k)T_{\lambda\rho}\left(-k\right)\rangle^{IR} _{even}  =  \frac{i |m|}{96\pi}
\left[ \frac{1}{2}\left( \left(k_\mu k_\lambda \eta_{\nu\rho} + \lambda \leftrightarrow \rho \right) + \mu \leftrightarrow \nu \right) - \right. \\
 \left.\phantom{\frac{1}{2}} - \left( k_\mu k_\nu \eta_{\lambda\rho} + k_\lambda k_\rho \eta_{\mu\nu}  \right)
 - \frac{k^2}2 \left( \eta_{\mu\lambda} \eta_{\nu\rho} + \eta_{\mu\rho} \eta_{\nu\lambda} \right) + k^2 \eta_{\mu\nu}\eta_{\lambda\rho} \right].
\end{multline}
The expression (\ref{IRlimitTT}) is transverse but not traceless because $\tau_g + \tau'_g \stackrel{IR}{\neq} 0$. 
To have a well-behaved IR limit we may add local counterterms to cancel the whole IR expression. That may be accomplished by simply performing the shifts
\be
\tau_g \rightarrow \tau_g - \frac{i}{24\pi} \left|\frac{m}{k}\right|, \quad \tau'_g \rightarrow \tau'_g + \frac{i}{48\pi} \left|\frac{m}{k}\right|.
\ee
These shifts correspond to the addition of a set of local counterterms in the expression (\ref{TTAfterCT}) and 
they do not change the UV behavior since they go to zero in that limit.

\subsection{Two-point function of the spin 3 current}
\label{ssec:J3J3}

Let us recall that we have postulated for the spin 3 current an action term of the form 
\begin{equation} \label{SintB} 
S_{int} \sim \int d^3x\, J_{\mu\nu\lambda} b^{\mu\nu\lambda},
\end{equation}
where $b$ is a completely symmetric 3rd order tensor (in this subsection we
assume $h_{\mu\nu}=0$ for simplicity). This interaction term gives rise to the
following b-field-fermion-fermion vertex $V_{bff}$
\begin{equation} \label{vertex3jff}
\frac 12  \left(\gamma_{(\mu_1} q_{2\mu_2} q_{2\mu_3)}+ q_{1(\mu_1}
q_{1\mu_2}\gamma_{\mu_3) }\right) 
-\frac 53 q_{1(\mu_1} \gamma_{\mu_2} q_{2\mu_3)}+ \frac 13
\eta_{(\mu_1\mu_2}\gamma_{\mu_3)} \left(q_1\!\cdot\!q_2 +m^2\right),
\end{equation}
where $q_1$ and $q_2$ are the incoming momenta of the two fermions. For a spin $n$ current, the analogous vertex can be obtained from the formula
\begin{equation}
V_{bff}: \langle \gamma z\rangle \, e^{i\langle(q_1-q_2)z\rangle} f\left(-2
\langle q_1q_2\rangle\langle z z\rangle
+4\langle q_1 z\rangle\langle q_2 z\rangle-2 m^2 \langle z
z\rangle\right)\label{VBff}
\end{equation}
by differentiating with respect to $z$ the right number of times (and setting
$z=0$).

As usual the contribution from the 2-point function comes from the bubble diagram
with incoming and outgoing momentum $k_\mu$. 
Using the  $V_{bff}$ vertex the bubble diagram gives 
\begin{eqnarray}\label{Tmmmnnn}
 \tilde J_{\mu_1\mu_2\mu_3\nu_1\nu_2\nu_3}(k)& =& \int\frac
{d^3p}{(2\pi)^3}\, {\rm Tr} \left( \frac 1{\slashed{p}-m} \left[ \frac 12 
\left(\gamma_{(\nu_1} (p-k)_{\nu_2} (p-k)_{\nu_3)} 
 + p_{(\nu_1} p_{\nu_2}\gamma_{\nu_3) }\right) \right.\right.\0\\
&&\left.+\frac 53 p_{(\nu_1} \gamma_{\nu_2} (p-k)_{\nu_3)}
- \frac 13 \eta_{(\nu_1\nu_2}\gamma_{\nu_3)} \left(p\!\cdot\!(p-k)
-m^2\right)\right]
 \frac 1{\slashed{p}-\slashed{k}-m} \0\\
&&\cdot \left[ \frac 12  \left(\gamma_{(\mu_1} (p-k)_{\mu_2} (p-k)_{\mu_3)} 
 + p_{(\mu_1} p_{\mu_2}\gamma_{\mu_3) }\right) \right.\0\\
&&\left.\left.+\frac 53 p_{(\mu_1} \gamma_{\mu_2} (p-k)_{\mu_3)}
-\frac 13 \eta_{(\mu_1\mu_2}\gamma_{\mu_3)} \left(p\!\cdot\!(p-k)
-m^2\right)\right]\right).
\end{eqnarray}
The parity-even part of the final result is given by
\be
\begin{split} \label{J3J3FullEven}
\tilde J^{(even)}_{\mu_1\mu_2\mu_3\nu_1\nu_2\nu_3}(k) & =  \tau_b \left(\frac{k^2}{m^2}\right) |k|^5 \pi_{\mu_1 \mu_2}  \pi_{\mu_3 \nu_1} \pi_{\nu_2 \nu_3} 
+ \tau'_b \left(\frac{k^2}{m^2}\right) |k|^5 \pi_{\mu_1 \nu_1}  \pi_{\mu_2 \nu_2} \pi_{\mu_3 \nu_3}\\
& + \mathcal{A}^{(even)}_{\mu_1\mu_2\mu_3\nu_1\nu_2\nu_3},
\end{split}
\ee
where
\begin{multline}
\tau_b  =  \frac{i}{288\pi k^6} \left[  6|k| |m| \left( k^4 +8k^2 m^2 -32m^4 \right) - \phantom{\frac{|k|}{2|m|}} \right.  \\
\left. - \left( 3 \left(k^2-4 m^2\right)^3 + 8 m^2 \left(k^2-6 m^2\right) \left(k^2+4 m^2\right) \right) \ArcTanh{\frac{|k|}{2|m|}} \right],
\end{multline}
\begin{multline}
\tau'_b =  \frac{i}{216\pi k^6}\left[  -6|k| |m| \left( k^4 - \frac{8}{3}k^2 m^2 +16m^4 \right) + \phantom{\frac{|k|}{2|m|}} \right. \\
\left.  + 3\left( k^2 - 4m^2 \right)^2 \left( k^2 + 4m^2 \right)\ArcTanh{\frac{|k|}{2|m|}} \right]
\end{multline}
and $\mathcal{A}^{(even)}$ corresponds to a set of contact terms that are not transverse but may be subtracted by local counterterms. It is given by
\begin{multline}
\mathcal{A}^{(even)}_{\mu_{1}\mu_{2}\mu_{3}\nu_{1}\nu_{2}\nu_{3}} = 
\frac{i m^3}{9\pi}\left[ \frac{3}{4} k_{\mu_1}k_{\nu_1}\eta_{\mu_2\mu_3}\eta_{\nu_2\nu_3}+\frac{7}{8}\left( k_{\mu_1}k_{\mu_2} \eta_{\nu_1\nu_2}\eta_{\mu_3\nu_3}
 + k_{\nu_1}k_{\nu_2} \eta_{\mu_1\mu_2}\eta_{\mu_3\nu_3} \right)   \right. \\
\left. + \frac{32}{15}m^2 \eta_{\mu_1\nu_1}\eta_{\mu_2\nu_2}\eta_{\mu_3\nu_3} 
+ \frac{52}{15}m^2 \eta_{\mu_1\nu_1}\eta_{\mu_2\mu_3}\eta_{\nu_2\nu_3} - \frac{3}{4}k^2\eta_{\mu_1\nu_1}\eta_{\mu_2\mu_3}\eta_{\nu_2\nu_3}\right].
\end{multline}
The parity-odd part is given by
\be
\begin{split}  \label{J3J3FullOdd}
\tilde J^{(odd)}_{\mu_1\mu_2\mu_3\nu_1\nu_2\nu_3}(k) & =  k^4 \epsilon_{\mu_1 \nu_1 \sigma}k^\sigma\left[ \kappa_b \left(\frac{k^2}{m^2}\right) 
 \pi_{\mu_2 \mu_3} \pi_{\nu_2 \nu_3} + \kappa'_b \left(\frac{k^2}{m^2}\right) \pi_{\mu_2 \nu_2} \pi_{\mu_3 \nu_3} \right] \\ 
& + \mathcal{A}^{(odd)}_{\mu_1\mu_2\mu_3\nu_1\nu_2\nu_3},
\end{split}
\ee
where
\begin{eqnarray}
\kappa_b & = & \frac{m}{72\pi |k|^5} \left[  -20|k|^3 |m| + 16|k| |m|^3  + \left( k^4 - 32m^4 \right)\ArcTanh{\frac{|k|}{2|m|}} \right],  \\
\kappa'_b & = & \frac{m}{18\pi |k|^5}\left[  2|k|^3 |m| + 8|k| |m|^3 - \left( k^2 - 4m^2 \right)^2 \ArcTanh{\frac{|k|}{2|m|}} \right],
\end{eqnarray}
and, as before, $\mathcal{A}^{(odd)}$ corresponds to a set of contact terms that are not transverse but may be subtracted by local counterterms. 
It is given by
\begin{multline}
\mathcal{A}^{(odd)}_{\mu_{1}\mu_{2}\mu_{3}\nu_{1}\nu_{2}\nu_{3}} = 
-\frac{\mathrm{sign}(m)|m|^2}{16\pi}\epsilon_{\mu_1\nu_1\sigma}k^\sigma\left[ \left(k_{\mu_2}k_{\mu_3}\eta_{\nu_2\nu_3} 
+ k_{\nu_2}k_{\nu_3}\eta_{\mu_2\mu_3}\right) + \frac{128}{27}m^2 \eta_{\mu_2\nu_2}\eta_{\mu_3\nu_3} \right.\\
\left. +\frac{32}{27}m^2 \eta_{\mu_2\mu_3}\eta_{\nu_2\nu_3} - k^2 \eta_{\mu_2\mu_3}\eta_{\nu_2\nu_3}  \right].
\end{multline}

\subsubsection{Even parity UV and IR limits}

In the UV limit, i.e. $\left| \frac{m}{k} \right| \to 0$, we find
\be
\lim_{\left| \frac{m}{k} \right| \to 0}\tau_b = -\frac{3}{4} \lim_{\left| \frac{m}{k} \right| \to 0}\tau'_b = \frac{1}{192}.
\ee
In the IR limit, i.e. $\left| \frac{k}{m} \right| \to 0$, we find
\be
\tau_b = \frac{8}{135\pi}\left| \frac{m}{k} \right| + \mathcal{O}\left( \left| \frac{k}{m} \right| \right) ,\quad \tau'_b 
= -\frac{4}{135\pi}\left|\frac{m}{k}\right|+ \mathcal{O}\left( \left| \frac{k}{m} \right| \right).
\ee
As in the case of the IR limit of the 2-point function of the stress-energy tensor, these leading divergent contributions of 
the form factors give rise to a set of contact terms in the IR that are all proportional to $|m|$. To add counter terms to 
make the IR well-behaved is equivalent to perform the following shift in the form factors $\tau_b$ and $\tau'_b$:
\be
\tau_b \rightarrow \tau_b - \frac{8 i}{135\pi}\left| \frac{m}{k} \right|,\quad \tau'_b \rightarrow \tau'_b + \frac{4 i}{135\pi}\left|\frac{m}{k}\right|.
\ee

\subsubsection{Odd parity UV}

In the UV limit we find
\be
\kappa_b = \frac{1}{144}\frac{m}{|k|} + \mathcal{O}\left( \left|\frac{m}{k}\right|^2 \right), \quad \kappa'_b = -\frac{1}{36}\frac{m}{|k|} 
+ \mathcal{O}\left( \left|\frac{m}{k}\right|^2 \right).
\ee
As in the previous cases the UV is specified by the leading term in $\frac
m{|k|}$. We get (after Wick rotation)
\begin{eqnarray} \label{TmmmnnnoddUV}
&& \tilde J^{(odd,UV)}_{\mu_1\mu_2\mu_3\nu_1\nu_2\nu_3}(k)=\frac{1}{4}\frac 
m{|k|}\,
\epsilon_{\mu_1\nu_1 \sigma} k^\sigma\Big{[} \frac 1{12} k_{\mu_2}k_{\mu_3}
k_{\nu_2}k_{\nu_3}-\frac 29 k^2 k_{\mu_3} k_{\nu_3}\eta_{\mu_2\nu_2} \0\\
&&+\frac {k^2}{36}\left( k_{\nu_2}k_{\nu_3} \eta_{\mu_2\mu_3}+
k_{\mu_2}k_{\mu_3} \eta_{\nu_2\nu_3}\right)
+\frac 19 k^4 \eta_{\mu_2\nu_2} \eta_{\mu_3\nu_3}-\frac 1{36}k^4
\eta_{\mu_2\mu_3} \eta_{\nu_2\nu_3}\Big{]}.
\end{eqnarray}
From now on in this section we understand symmetrization among $\mu_1,\mu_2,\mu_3$ and among $\nu_1,\nu_2,\nu_3$. 
The anti-Wick rotation does not yield any change. We can contract (\ref{TmmmnnnoddUV}) with any $k^{\mu_i}$ and 
any two indexes $\mu_i$ and find zero. Therefore (\ref{TmmmnnnoddUV}) is conserved and traceless (it satisfies eq.(\ref{trace0cond})). 

We have obtained the same result (\ref{TmmmnnnoddUV}) with the method illustrated in
Appendix \ref{sec:Lauricella}.

\subsubsection{Odd parity IR}

In the IR limit we find
\be
\kappa_b = \frac{8}{27\pi}\frac{m^2}{k^2}  + \frac{1}{240\pi} + \mathcal{O}\left( \left|\frac{k}{m}\right|^2 \right), \quad \kappa'_b 
= - \frac{8}{27\pi}\frac{m^2}{k^2} - \frac{2}{135\pi} + \mathcal{O}\left( \left|\frac{m}{k}\right|^2 \right).
\ee
Once again the IR limit contain divergent contributions that can be treated by adding local counter terms, which is equivalent to 
perform the following shifts on the form factors:
\be
\kappa_b \rightarrow \kappa_b + \frac{8}{27\pi} \frac{m^2}{k^2},\quad \kappa'_b \rightarrow \kappa'_b -\frac{8}{27\pi} \frac{m^2}{k^2}.
\ee
The final result in Lorentzian metric (obtained with the two different methods
above) is
\begin{multline}\label{TmmmnnnIR}
\tilde J^{(odd,IR)}_{\mu_1\mu_2\mu_3\nu_1\nu_2\nu_3}(k)=\frac 1{4\pi}
\epsilon_{\mu_1\nu_1 \sigma} k^\sigma\Big{[}  \frac{1}{60}k^4
\eta_{\mu_2\mu_3} \eta_{\nu_2\nu_3} -  \frac{8}{135}k^4\eta_{\mu_2\nu_2} \eta_{\mu_3\nu_3}\\
-\frac{1}{60} k^2\left( k_{\nu_2}k_{\nu_3} \eta_{\mu_2\mu_3}+ k_{\mu_2}k_{\mu_3} \eta_{\nu_2\nu_3}\right) +
 \frac{16}{135} k^2 k_{\mu_2} k_{\nu_2}\eta_{\mu_3\nu_3} - \frac{23}{540}  k_{\mu_2}k_{\mu_3} k_{\nu_2}k_{\nu_3}\Big{]}.
\end{multline}
 
The trace of (\ref{TmmmnnnIR}) does not vanish. However at this point we must avoid a semantic trap.
A nonvanishing trace of this kind does not contradict the fact that it represents a fixed point
of the renormalization group. An RG fixed point is expected to be conformal, but this
means vanishing of the e.m. trace, not necessarily of the trace of the spin three current.

\subsubsection{The lowest order effective action for the field $B$}

The odd 2-point correlator in a scaling UV limit similar to (\ref{scalinglimit}), Fourier anti-transformed and inserted in (\ref{Wap}), gives rise to the action term
\begin{eqnarray} \label{BUVaction}
S^{(UV)}\sim \int d^3x\!\! &&\!\epsilon_{\mu_1\nu_1\sigma} \Big{[}
3\partial^\sigma B^{\mu_1\mu_2\mu_3} 
\partial_{\mu_2} \partial_{\mu_3} \partial_{\nu_2} \partial_{\nu_3} 
B^{\nu_1\nu_2\nu_3}-8 \partial^\sigma B^{\mu_1\mu_2\mu_3}\square 
\partial_{\mu_3}\partial_{\nu_3} B^{\nu_1\nu_3}{}_{\mu_2}\0\\
&&+2 \partial^\sigma B^{\mu_1\lambda}{}_\lambda \square  \partial_{\nu_2}
\partial_{\nu_3} B^{\nu_1\nu_2\nu_3} 
+4\partial^\sigma B^{\mu_1\mu_2\mu_3} \square^2 B^{\nu_1}{}_{\mu_2\mu_3}\0\\
&&-
\partial^\sigma B^{\mu_1\lambda}{}_\lambda \square^2 B^{\nu_1\rho}{}_\rho
\Big{]},
\end{eqnarray}
where $B_{\mu\nu\lambda}=b_{\mu\nu\lambda}+\ldots$.
This is the lowest order term of the analog of the CS action for the field $B$. 
This theory is extremely constrained. The field $B$ has 10 independent
components. The gauge freedom
counts 6 independent functions, the conservation equations are 3. The generalized Weyl (g-Weyl) invariance
implies two additional degrees of freedom. So altogether 
the constraints are more than the degrees of freedom. The question is
whether such CS actions contain nontrivial (i.e. non pure gauge) solutions.

In a similar way (\ref{TmmmnnnIR}) gives rise to the action
\begin{eqnarray} \label{BIRaction}
S^{(IR)}= \frac 1{32\pi} \frac 1{540} \int d^3x\!\! &&\!\epsilon_{\mu_1\nu_1\sigma} \Big{[}
-23\partial^\sigma B^{\mu_1\mu_2\mu_3}\partial_{\mu_2} \partial_{\mu_3} \partial_{\nu_2} \partial_{\nu_3} B^{\nu_1\nu_2\nu_3}\0\\
&&+64\partial^\sigma B^{\mu_1\mu_2\mu_3}\square 
\partial_{\mu_3}\partial_{\nu_3} B^{\nu_1\nu_3}{}_{\mu_2} -
18 \partial^\sigma B^{\mu_1\lambda}{}_\lambda \square  \partial_{\nu_2}\partial_{\nu_3} B^{\nu_1\nu_2\nu_3} \0\\
&&-32\partial^\sigma B^{\mu_1\mu_2\mu_3} \square^2 B^{\nu_1}{}_{\mu_2\mu_3}+9\partial^\sigma B^{\mu_1\lambda}{}_\lambda \square^2 B^{\nu_1\rho}{}_\rho
\Big{]}.
\end{eqnarray}
This action is invariant under (\ref{Bgaugetransf}), but not under (\ref{Bweyltransf}).

{\bf Remark}  The action (\ref{BUVaction}) is similar to eq.(30) of \cite{pope}.
The latter is written in terms of spinor labels, therefore the relation is not
immediately evident. After turning to the ordinary notation, eq.(30) of
\cite{pope} becomes
\begin{eqnarray}
\sim \int d^3x\!\! &&\!\epsilon_{\mu_1\nu_1\sigma} \Big{[}\frac 32
\partial^\sigma h^{\mu_1\mu_2\mu_3} 
\partial_{\mu_2} \partial_{\mu_3} \partial_{\nu_2} \partial_{\nu_3} 
h^{\nu_1\nu_2\nu_3}-4 \partial^\sigma h^{\mu_1\mu_2\mu_3}\square 
\partial_{\mu_3}\partial_{\nu_3} h^{\nu_1\nu_3}{}_{\mu_2}\0\\
&& 
+2\partial^\sigma h^{\mu_1\mu_2\mu_3} \square^2 h^{\nu_1}{}_{\mu_2\mu_3}
\Big{]}\label{Baction1}
\end{eqnarray}
and one can see that they are equal if we set $B^{\mu\lambda}{}_\lambda=0$ in (\ref{BUVaction}). 
The reason of the difference is that in \cite{pope} the field $h^{\mu\nu\lambda}$ is traceless, 
while in (\ref{BUVaction}) the field $B_{\mu\nu\lambda}$ is not. The presence of the trace part modifies the conservation law and thus the action.

\section{Chern-Simons effective actions}

In the previous section we have seen that the odd parity 2-point correlators of the
massive fermion model,
either in the IR or UV limit, are local and give rise to action terms which
coincide with the lowest
(second) order of the gauge CS action and gravity CS action for the 2-point function
of the gauge current and the e.m. tensor, respectively; and to the lowest order
of a CS-like action for the rank 3 tensor field $B$. It is natural to expect that
the n-th order terms of such CS actions will originate in a similar way from the
n-point functions of the relevant currents. In particular the next to leading
(third order) term  
in the CS actions is expected to be determined by the 3-point functions of the
relevant currents.
This is indeed so, but in a quite nontrivial way, with complications due both to
the regularization and to the way we take the IR and UV limit.

The purpose of this section is to elaborate on the properties of the gauge and
gravity CS actions, (\ref{gaugeCS}) and (\ref{gravityCS}), respectively, in
order to prepare the ground for the following discussion. The point we want to
stress here is that in order to harmonize the formalism with the perturbative
expansion in quantum field theory we need perturbative cohomology. The latter is
explained 
in detail in Appendix \ref{sec:pertcohomology}. It consists of a sequence of
coboundary operators which approximate the full cohomology: in the case of a
gauge theory the sequence reduces to two elements, in the case of gravity or
higher tensor theories the sequence is infinite.

\subsection{CS term for the gauge field}

Let us start with the gauge case. The action (\ref{gaugeCS}) splits into two
parts, $CS =CS^{(2)}+ CS^{(3)}$,  of order two and three, respectively, in the
gauge field $A$. The second term is expected to come from the 3-point function of
the gauge current. Gauge invariance splits as follows
\be \label{pertgaugeinv}
\delta^{(0)} CS^{(2)}=0,\quad\quad \delta^{(1)} CS^{(2)}+\delta^{(0)}
CS^{(3)}=0.
\ee
These equations reflect themselves in the conservation laws, which also split 
into two equations. The conservation law for the 2-point function is simply 
the vanishing of the divergence (on any index) of the latter, while for the 3-point 
function it does not consist in the vanishing of the divergence of the latter,
 but involves also contributions from 2-point functions. More precisely 
\be \label{3rdordergaugeWI}
&&\partial_x^\mu \langle 0|\mathcal{T} J_\mu^a(x)
J_\nu^b(y)J_\lambda^c(z)|0\rangle\0\\
&& = i
f^{abc'} \delta(x-y) \langle 0|\mathcal{T} J_\nu^{c'}(x) 
J_\lambda^c(z)|0\rangle+
f^{acc'}  \delta(x-z)\langle 0|\mathcal{T} J_\lambda^{c'}(x) J_\nu^b(y)|0\rangle,
\ee
which in momentum space becomes
\be \label{3ptconserv}
-i q^\mu \tilde J_{\mu\nu\lambda}^{abc}(k_1,k_2) +f^{abc'} \tilde
J_{\nu\lambda}^{c'c}(k_2)+ 
f^{acc'} \tilde J_{\lambda\nu}^{c'b}(k_1)=0,
\ee
where $q=k_1+k_2$ and $\tilde J_{\mu\nu}^{ab}(k)$ and $\tilde
J_{\mu\nu\lambda}^{abc}(k_1,k_2)$ are Fourier transform of the 2- and 3-point
functions,
respectively.

\subsection{Gravitational CS term}

Let us consider next the gravitational CS case. Much as in the previous case we
split the action 
(\ref{gravityCS}) in pieces according to the number of $h_{\mu\nu}$ contained in
them. This time however the number of pieces is infinite:
\begin{eqnarray} \label{CS23}
CS_g= \kappa 
\int d^3x\, \epsilon^{\mu\nu\lambda}
\left(\partial_\mu\omega_\nu^{ab}
\omega_{\lambda b a}+ \frac 23 \omega_{\mu a}{}^b \omega_{\nu b}{}^c
\omega_{\lambda c}{}^a\right)= CS_g^{(2)}+CS_g^{(3)}+\ldots,
\end{eqnarray}
where 
\begin{equation}
CS_g^{(2)}=\frac{\kappa}2\int d^3x\,  \epsilon_{\sigma\nu\rho}\, h^{\lambda
\rho}\left( \partial^\sigma
\partial_\lambda \partial_b h^{b\nu}-\partial^\sigma \square h_\lambda^\nu
\right)\label{CS2h}
\end{equation}
and 
\begin{eqnarray}\label{CS3h}
CS_g^{(3)} &=&\frac {\kappa}4 \int d^3x \, \epsilon^{\mu\nu\lambda} \Big{(}2
\partial_a h_{\nu b}
\partial_{\lambda} h_\sigma^b \partial_\mu h^{\sigma a}
-2 \partial_a h_{\mu}^b \partial^c h_{b\nu} \partial^a h_{c\lambda}-\frac
23\partial_a h_{\mu}^b \partial_b h_\nu^c \partial_c h^a_{\lambda} \0\\
&& -2\partial _\mu \partial^b h^a_\nu (h_a^c \partial_c h_{b\lambda} -h_b^c
\partial_c h_{a\lambda}) 
+\partial _\mu \partial^b h^a_\nu 
(h_\lambda^c \partial_a h_{bc}
-\partial_a h_\lambda^c h_{bc})\0\\
&&+
\partial_\mu \partial^b h_\nu^a 
\left( \partial_b h_\lambda^c h_{ac}- h_\lambda^c\partial_b h_{ac}
\right) 
-h_\lambda^\rho h_\rho^a\partial_\mu \left(\square h_{a\nu}
-\partial_a\partial_bh^b_\nu\right)
\Big{)}.
\end{eqnarray}
Invariance of $CS_g$ under diffeomorphisms also splits into infinite many
relations. The first two, which are relevant to us here, are
\begin{equation} \label{deltaCSg}
\delta_\xi^{(1)} CS_g^{(0)}=0,\quad\quad \delta_\xi^{(1)} CS_g^{(0)}+
\delta_\xi^{(0)}CS_g^{(1)}=0,
\end{equation}
where $\xi$ is the parameter of diffeomorphisms. Similar relations hold also for
Weyl transformations.

Such splittings correspond to the splittings of the Ward identities for 
diffeomorphisms and Weyl transformations derived from the generating function
(\ref{Wap}).
The lowest order WI is just the vanishing of the divergence of the 2-point e.m.
tensor correlators.
The next to lowest order involves 2-point as well as 3-point functions of the e.m.
tensor:
\begin{eqnarray} \label{3rdorderWI}
&&\langle 0|\mathcal{T}\{\del^\mu T_{\mu\nu}
(x)T_{\lambda\rho}(y)T_{\alpha\beta}(z)\}|0\rangle\0\\
&&=i\left\{ 2\frac{\del}{\del x^{(\alpha}}\left[ \delta(x-z)\langle
0|\mathcal{T}\{T_{\beta)\nu}(x) T_{\lambda\rho}(y)\}|0\rangle\right] + 
2\frac{\del}{\del x^{(\lambda}}  \left[ \delta(x-y)\langle
0|\mathcal{T}\{T_{\rho)\nu}(x) T_{\alpha\beta}(z)\}|0\rangle\right] \right.\0\\
&&-\,\frac{\del}{\del x_\tau}\delta(x-z)\eta_{\alpha\beta} \langle
0|\mathcal{T}\{T_{\tau\nu}(x)T_{\lambda\rho}(y)\}|0\rangle -\,\frac
{\del}{\del x_\tau}\delta(x-y)\eta_{\lambda\rho} \langle
0|\mathcal{T}\{T_{\tau\nu}(x)T_{\alpha\beta}(z)\}|0\rangle\0\\
&&+ \left. \,\frac{\del}{\del{x^\nu}}
\delta(x-z)\langle
0|\mathcal{T}\{T_{\lambda\rho}(y)T_{\alpha\beta}(x)\}|0\rangle +
\,\frac{\del}{\del{x^\nu}}
\delta(x-y)\langle
0|\mathcal{T}\{T_{\lambda\rho}(x)T_{\alpha\beta}(z)\}|0\rangle\right\}.
\end{eqnarray}
In momentum space, denoting by $\tilde T_{\mu\nu\lambda\rho}(k)$ and by 
$\tilde T_{\mu\nu\lambda\rho\alpha\beta}(k_1,k_2)$ the 2-point and 3-point function,
respectively, this formula becomes
\begin{eqnarray} \label{3ptgravityWI}
i q^\mu  \tilde T_{\mu\nu\lambda\rho\alpha\beta}(k_1,k_2)&=& 2q_{(\alpha} \tilde
T_{\beta)\nu\lambda\rho}(k_1)+2
q_{(\lambda}\tilde T_{\rho)\nu\alpha\beta}(k_2)- \eta_{\alpha\beta} k_2^\tau
\tilde T_{\tau\nu\lambda\rho}(k_1)\0\\
&&-\eta_{\lambda\rho} k_1^\tau \tilde T_{\tau\nu\alpha\beta}(k_2)+k_{2\nu}
\tilde T_{\alpha\beta\lambda\rho}(k_1)+k_{1\nu}
\tilde T_{\lambda\rho\alpha\beta}(k_2),
\end{eqnarray}
where round brackets denote symmetrization normalized to 1.
  
From the action term (\ref{CS3h}), by differentiating three times with respect
to
$h_{\mu\nu}(x)$,$h_{\lambda\rho}(y)$ and $h_{\alpha\beta}(z)$ and 
Fourier-transforming the result one gets a sum of local terms in momentum space
(see Appendix \ref{sec:3rdorderCS}),
to be compared with the IR and UV limit of the 3-point e.m. tensor correlator.

\subsection{CS term for the $B$ field}
\label{ssec:s3cs}

Here we would like to understand the nature of the ``CS-like'' terms obtained in the IR 
and UV limits, and especially to understand how it is possible that they are different in 
the spin-3 case, unlike what we saw in spin-1 and spin-2 \footnote{In the literature one can find two kinds 
of generalizations of the CS action in 3d to higher spins (for a general review on higher spin theories, 
see \cite{solvay2004,sorokin}). One leads to quadratic equations of motion, 
the other to higher derivative equations of motion. The first kind of theories are nicely summarized in \cite{campoleoni}. The second kind of theories, to our best knowledge, was introduced in \cite{pope} (following \cite{blencowe}). This splitting was already shadowed in \cite{WittenCS}.}. For this purpose, we use a higher-spin ``geometric'' construction originally developed in \cite{deWit:1979sib}. In the spin-3 case the linearised ``Christoffel connection'' is given by the so-called second affinity defined by
\begin{eqnarray} \label{s3lcc}
\Gamma_{\alpha_1 \alpha_2 ; \beta_1 \beta_2 \beta_3} &=& \frac{1}{3} \left\{
\partial_{\alpha_1} \partial_{\alpha_2} 
B_{\beta_1 \beta_2 \beta_3} - \frac{1}{2} \left( \partial_{\alpha_1}
\partial_{\beta_1} B_{\alpha_2 \beta_2 \beta_3} + 
\partial_{\alpha_1} \partial_{\beta_2} B_{\alpha_2 \beta_1 \beta_3} \right.
\right. 
\nonumber \\
&& \left. + \partial_{\alpha_1} \partial_{\beta_3} B_{\alpha_2 \beta_1 \beta_2}
 + \partial_{\alpha_2} \partial_{\beta_1} B_{\alpha_1 \beta_2 \beta_3} +
\partial_{\alpha_2} \partial_{\beta_2} B_{\alpha_1 \beta_1 \beta_3} +
\partial_{\alpha_2} \partial_{\beta_3} B_{\alpha_1 \beta_1 \beta_2} \right)
\nonumber \\
&& + \partial_{\beta_1} \partial_{\beta_2} B_{\alpha_1 \alpha_2 \beta_3}
 + \partial_{\beta_1} \partial_{\beta_3} B_{\alpha_1 \alpha_2 \beta_2} +
\partial_{\beta_2} \partial_{\beta_3} B_{\alpha_1 \alpha_2 \beta_1} \bigg\}.
\end{eqnarray}
Under the gauge transformation (\ref{Bgaugetransf}) this ``connection'' transforms as
\begin{equation} \label{s3LCgt}
\delta_\Lambda \Gamma_{\alpha \beta ; \mu\nu\rho}
 = \partial_\mu \partial_\nu \partial_\rho \Lambda_{\alpha\beta}.
\end{equation}

The natural generalisation of (the quadratic part of) the spin-1 and the spin-2 CS action
term to the spin-3 case is given by
\begin{eqnarray} \label{IcsB}
I_{\mathrm{CS}}[B] &\equiv& \int d^3 x\, \epsilon^{\mu\sigma\nu}\,
\Gamma^{\alpha \beta}{}_{;\mu \rho \lambda}\, \partial_\sigma
 \Gamma^{\rho \lambda}{}_{;\nu \alpha \beta}
\nonumber \\
 &=& \frac{1}{3} \int d^3 x\, \epsilon_{\mu\sigma\nu} \big( \partial_\alpha
\partial_\beta B^{\mu\alpha\beta}\, \partial^\sigma
  \partial_\rho \partial_\lambda B^{\nu\rho\lambda} + 2\, \partial_\alpha \Box
B^{\mu\alpha\beta}\, \partial^\sigma \partial_\rho
  B^{\nu\rho}{}_\beta
\nonumber \\
 && \qquad\qquad\qquad +\: \Box B^{\mu\alpha\beta}\, \partial^\sigma \Box
B^{\nu}{}_{\alpha\beta} \big) + \; \textrm{(boundary terms)}.
\end{eqnarray}
From (\ref{s3LCgt}) directly follows that this CS term is gauge invariant (up to boundary terms).
In the spin-3 case one can construct another 5-derivative CS term by using Fronsdal tensor (or spin-3 
``Ricci tensor'') defined by
\begin{eqnarray}
\mathcal{R}_{\mu\nu\rho} &\equiv& \Gamma^\alpha{}_{\alpha ; \mu\nu\rho}
\nonumber \\ 
&=& \frac{1}{3} \big\{ \Box B_{\mu\nu\rho}
 - \partial^\alpha ( \partial_\mu B_{\alpha\nu\rho} 
 + \partial_\nu B_{\alpha\rho\mu} + \partial_\rho B_{\alpha\mu\nu} )
\nonumber \\
&& \quad\; +\: \partial^\mu \partial^\nu B_{\rho\alpha}{}^\alpha 
 + \partial^\rho \partial^\mu B_{\nu\alpha}{}^\alpha + \partial^\nu
\partial^\rho B_{\mu\alpha}{}^\alpha  \big\}.
\end{eqnarray}
Using this tensor one can defined another CS action term
\begin{eqnarray} \label{IcsFB}
I'_{\mathrm{CS}}[B] &\equiv& \int d^3 x\, \epsilon^{\mu\sigma\nu}\,
\mathcal{R}_{\mu \rho \lambda}\, \partial_\sigma
 \mathcal{R}_{\nu}{}^{\rho \lambda}
\nonumber \\
 &=& \frac{1}{9} \int d^3 x\, \epsilon_{\mu\sigma\nu} \big( 2\, \partial_\alpha
\partial_\beta B^{\mu\alpha\beta}\, \partial^\sigma
  \partial_\rho \partial_\lambda B^{\nu\rho\lambda} + 2\, \partial_\alpha \Box
B^{\mu\alpha\beta}\, \partial^\sigma \partial_\rho
  B^{\nu\rho}{}_\beta
\nonumber \\
 && \qquad\qquad\qquad -\: 2\, \partial_\alpha \partial_\beta
B^{\mu\alpha\beta}\, \partial^\sigma \Box B^{\nu\rho}{}_{\rho}
 + \Box B^{\mu\alpha\beta}\, \partial^\sigma \Box B^{\nu}{}_{\alpha\beta}
\nonumber \\
 && \qquad\qquad\qquad +\:  \Box B^{\mu\alpha}{}_\alpha\, \partial^\sigma \Box
B^{\nu\rho}{}_\rho \big) + \; \textrm{(boundary terms)}.
\end{eqnarray}
The presence of two CS terms in the spin-3 case explains why there is a priori no reason to expect from UV and IR limits to lead to the same CS-like term.

Now it is easy to see that the following combination
\begin{equation} \label{csUVcomb}
5\,I_{\mathrm{CS}}[B] - 3\,I'_{\mathrm{CS}}[B]
\end{equation}
exactly gives the effective action term (\ref{BUVaction}) which we obtained from the one-loop calculation.

To understand why the combination (\ref{csUVcomb}) represents a generalization of the spin-2 CS term 
(gravitational CS term), one has to take into account the symmetry under generalized Weyl (g-Weyl) transformations, 
which for spin-3 is given by (\ref{Bweyltransf}). It can be shown that the CS terms 
(\ref{IcsB}) and (\ref{IcsFB}) are not g-Weyl-invariant, but that (\ref{csUVcomb}) is the \emph{unique g-Weyl-invariant} linear combination thereof.

It is then not surprising that the effective current $J_{\mu\nu\rho}$ obtained from (\ref{csUVcomb}) is
 proportional to the spin-3 ``Cotton tensor'' studied in \cite{Damour:1987vm}. It is the gauge- and g-Weyl- invariant 
conserved traceless totally symmetric tensor with the property that if it vanishes then the gauge field is g-Weyl-trivial. 
With this we have completed the demonstration that on the linear level the spin-3 CS term is a natural generalisation of the spin-2 CS term.

For completeness we add that the combination
\be
\frac 1{192\pi} \left( -\frac {41}3 I_{\mathrm{CS}}[B]+ 9  I'_{\mathrm{CS}}[B]\right) \label{csIRcomb}
\ee
reproduces (\ref{BIRaction}), which is not g-Weyl invariant. 

\section{Three-point gauge current correlator: odd parity part}
\label{sec:3ptcurrent}

In this section we explicitly compute the 3-point current correlator of the current
$J^a_\mu(x)$.
The 3-point correlator for currents is given by the triangle diagram. The three
external
momenta are $q,k_1,k_2$. $q$ is incoming, while $k_1,k_2$ are outgoing and of
course momentum conservation implies
$q=k_1+k_2$. The relevant Feynman diagram is 
\be
\tilde J_{\mu\nu\lambda}^{1,abc}(k_1,k_2)\label{Jmnl}
 =i \int \frac{d^3p}{(2\pi)^3} {\rm
Tr}\left(\gamma_\mu T^a \frac 1{\slashed{p} -m }   \gamma_\nu T^b
\frac 1{\slashed{p}-\slashed{k}_1 -m }\gamma_\lambda T^c\frac
1{\slashed{p}-\slashed{q} -m }\right)
\ee
to which we have to add the cross graph contribution
$\tilde J_{\mu\nu\lambda}^{2,abc}(k_1,k_2)=\tilde
J_{\mu\lambda\nu}^{1,acb}(k_2,k_1)$.
From this we extract the odd parity part and perform the integrals. The general
method is discussed
in subsection \ref{ssec:3ptTtriangle}, here we limit ourselves to the results.
Such results 
have already been presented in \cite{BL}, but since they are important for the
forthcoming discussion
we summarize them below. For simplicity we set
$k_1^2=k_2^2=0$, so the total energy of the process is
$E=\sqrt{q^2}=\sqrt{2k_1\!\cdot\! k_2}$.

Near the IR fixed point we obtain a series expansion of the type
\be 
\tilde J_{\mu\nu\lambda}^{abc(odd)}(k_1,k_2) \approx i\frac {\rm
1}{32\pi} 
\sum_{n=0}^\infty \left(\frac {E}{m}\right)^{2n} f^{abc} \tilde
I_{\mu\nu\lambda}^{(2n)}(k_1,k_2)\label{IRodd1}
\ee
and, in particular, the relevant term in the IR is
\be \label{IRodd2}
I_{\mu\nu\lambda}^{(0)}(k_1,k_2)= -12 \epsilon_{\mu\nu\lambda}.
\ee
The first thing to check is conservation. The current (\ref{Jmu}) should be conserved also at the quantum level, because
no anomaly is expected in this case. It is evident that the contraction with
$q^\mu$ 
does not give a vanishing result, as we expect because we must include also the
contribution from the
2-point functions, (\ref{3rdordergaugeWI}). But even including such contributions we
get
 \be 
-\frac 3{8\pi} f^{abc} q^\mu \epsilon_{\mu\nu\lambda} + \frac 1{4\pi} 
f^{abc}  \epsilon_{\nu\lambda\sigma}k_2^\sigma +
 \frac 1{4\pi}  f^{abc}  \epsilon_{\nu\lambda\sigma}k_1^\sigma\neq
0.\label{noconserv1}
\ee
Conservation is violated by a local term. Thus we can recover it by adding to
$I_{\mu\nu\lambda}^{(0)}(k_1,k_2)$ a
term $4  \epsilon_{\mu\nu\lambda}$.  This corresponds to correcting the
effective action by adding a counterterm 
\be \label{counterIR1}
-2 \int dx  \epsilon^{\mu\nu\lambda} f^{abc}A_\mu^a A_\nu^b
A_\lambda^c.
\ee
Adding this to the result from the 2-point correlator we reconstruct the full CS
action (\ref{gaugeCS}). 

This breakdown of conservation is surprising, therefore it is
important to 
understand where it comes from. To this end we consider the full theory for
finite $m$. The contraction
of the 3-point correlator with $q^\mu$ is given by
\be 
q^\mu {\tilde J}^{abc}_{\mu\nu\lambda} (k_1,k_2)  = -i \!\int\! \frac
{d^3p}{(2\pi)^3}
{\rm Tr} \left( \slashed{q} T^a \frac 1{\slashed{p}-m} \gamma_\nu T^b \frac
1{\slashed{p}-\slashed{k}_1-m} 
\gamma_\lambda T^c \frac 1{\slashed{p}-\slashed{q}-m} \right).\label{qJ1}
\ee 
Replacing $\slashed{q} =( \slashed{p}-m)- ( \slashed{p}-\slashed{q}-m)$
considerably simplifies the 
calculation. The final result for the odd parity part (after adding the cross diagram contribution, $1\leftrightarrow 2, b\rightarrow c, \nu\leftrightarrow \lambda$) is 
 \be\label{qJ4}
q^\mu {\tilde J}^{abc}_{\mu\nu\lambda} (k_1,k_2)  =  &
 -\frac i{4\pi} f^{abc} \epsilon_{\lambda\nu \sigma}k_1^\sigma \,\frac
{2m}{k_1}\arccoth\left( \frac {2m}{k_1} \right)\0\\
&- \frac i{4\pi} f^{abc} \epsilon_{\lambda\nu \sigma}k_2^\sigma \,\frac
{2m}{k_2}\arccoth\left( \frac {2m}{k_2} \right).
\ee
Therefore, as far as the odd part is concerned, the 3-point conservation
(\ref{3ptconserv}) reads
\begin{multline}
 -i q^\mu \tilde J_{\mu\nu\lambda}^{(odd)abc}(k_1,k_2) +f^{abd} \tilde
J_{\nu\lambda}^{(odd)dc}(k_2)+ 
f^{acd} \tilde J_{\lambda\nu}^{(odd)db}(k_1)\label{qJ5}\\
 =- \frac 1{4\pi} f^{abc} \epsilon_{\lambda\nu \sigma}\left(k_1^\sigma \,\frac
{2m}{k_1}\arccoth \left( \frac {2m}{k_1} \right)+ k_2^\sigma \,\frac
{2m}{k_2}\arccoth\left( \frac {2m}{k_2} \right)\right)\\
+ \frac 1{4\pi} f^{abc} \epsilon_{\lambda\nu \sigma}\left(k_1^\sigma \,\frac
{2m}{k_1}\arccoth\left( \frac {2m}{k_1} \right)+ k_2^\sigma \,\frac
{2m}{k_2}\arccoth\left( \frac {2m}{k_2} \right)\right)=0.
\end{multline}
Thus conservation is secured for any value of the parameter $m$. The fact that
in the IR limit we find a violation of the conservation is an 
artifact of the procedure we have used (in particular of the limiting procedure)
and we have 
to remedy by subtracting  suitable counterterms from the effective action. 
These subtractions are to be understood as (part of) the definition of 
our regularization procedure.

Something similar can be done also for the UV limit. However in the UV limit the
resulting effective action has a vanishing coupling $\sim \frac mE$, unless we
consider an $N\to
\infty$ limit theory, as outlined above. In order to guarantee invariance under
large gauge transformations we have also to fine tune the limit in such a way
that the $\kappa$ coupling be an integer. But even in the UV we meet the
problem 
of invariance breaking.

We will meet the same problem below for the 3-point function of the e.m. tensor.

\section{Three-point e.m. correlator: odd parity part}
\label{sec:3ptem}

We go now to the explicit calculation of the 3-point e.m. tensor correlator.
The three-point function is given by three contributions, the
bubble diagram, the triangle diagram and the cross triangle diagram.
We will focus in the sequel only on the odd parity part.

\subsection{The bubble diagram: odd parity}
\label{ssec:3ptTbubble}

The bubble diagram is constructed with one $V_{gff}$ vertex and one $V_{ggff}$
vertex. It
has an incoming line with momentum $q=k_1+k_2$ with Lorentz indices $\mu,\nu$,
and two outgoing lines have
momenta $k_1,k_2$ with Lorentz labels $\lambda,\rho$ and $\alpha,\beta$,
respectively. The internal running momentum is denoted
by $p$. The corresponding contribution is 
\begin{multline} 
D_{\lambda\rho\alpha\beta\mu\nu}(k_1,k_2) \label{Dk1k2}\\
= \frac i{128} \int \frac {d^3p}{(2\pi)^3}{\rm Tr} \left[ \frac 1
{{\slashed{p}}-m}t_{\lambda\rho\alpha
\beta\sigma} (k_2-k_1)^\sigma 
 \frac 1 {{\slashed{p}}-\slashed{q}-m}((2p_\mu -q_\mu)\gamma_\nu + \mu
\leftrightarrow \nu)\right]
\end{multline}
where
\begin{eqnarray}  \label{tlrabs}
t_{\lambda\rho\alpha\beta\sigma}= \eta_{\lambda\alpha}
\epsilon_{\rho\beta\sigma}+ \eta_{\lambda\beta} \epsilon_{\rho\alpha\sigma}   +
\eta_{\rho\alpha} \epsilon_{\lambda\beta\sigma}+ \eta_{\rho\beta}
\epsilon_{\lambda\alpha\sigma}. 
\end{eqnarray}
The odd part gives (the metric is Lorentzian)
\begin{eqnarray} 
\tilde D_{\lambda\rho\alpha\beta\mu\nu}(k_1,k_2)&=& \frac m{256\pi}
t_{\lambda\rho\alpha\beta\sigma}(k_2-k_1)^\sigma
\left( \eta_{\mu\nu} \left( 2m -\frac {q^2-4m^2}{|q|} \arctanh \frac
{|q|}{2m}\right)\right. \label{Dk1k2b}\nonumber\\
&& \left.+q_\mu q_\nu \left(  \frac {2m}{q^2} +\frac {q^2-4m^2}{|q|^3} \arctanh
\frac {|q|}{2m}\right)\right).
\end{eqnarray}
Saturating with $q^\mu$ we get
\be \label{qD}
 q^\mu\tilde D_{\lambda\rho\alpha\beta\mu\nu}(k_1,k_2)= \frac {m^2}{256\pi}
t_{\lambda\rho\alpha\beta\sigma}(k_2-k_1)^\sigma 2q_\nu.
\ee
This corresponds to an anomaly 
\begin{eqnarray}
{\cal A}_\xi \sim \int d^3x\, \partial_\nu \xi^\nu \epsilon_{\rho\beta\sigma}
h^{\lambda\rho} \partial^\sigma h_\lambda^\beta\label{Dxi}
\end{eqnarray}
which we have to subtract. This gives
\begin{multline} \label{Dk1k2c}
\tilde D_{\lambda\rho\alpha\beta\mu\nu}(k_1,k_2)\\
= \frac 1{256\pi} t_{\lambda\rho\alpha\beta\sigma}(k_2-k_1)^\sigma
\left( {q_\mu q_\nu}-\eta_{\mu\nu} q^2\right) \left(   \frac {2m^2}{q^2} +m
\frac {q^2-4m^2}{|q|^3} \arctanh \frac {|q|}{2m}\right).
\end{multline}
Taking the limit of the form factor (last round brackets) for $\frac m{|q|} \to
0$ (UV), we find 0 (the linear term in $\frac m{|q|}$ vanishes). 
Taking the limit $\frac m{|q|} \to \infty$ 
(IR) we find
\begin{equation} \label{Dk1k2d}
 \tilde D^{(IR)}_{\lambda\rho\alpha\beta\mu\nu}(k_1,k_2)= \frac 23\, \frac{1}{256\pi} 
t_{\lambda\rho\alpha\beta\sigma}(k_2-k_1)^\sigma\left( {q_\mu
q_\nu}-\eta_{\mu\nu} q^2\right).
\end{equation}
This corresponds to the action term
\begin{equation} \label{actiontermD}
 \sim \int d^3x\,(\square h - \partial_\mu \partial_\nu 
h^{\mu\nu})\,t_{\lambda\rho\alpha\beta\sigma}\, \left(h^{\lambda\rho} 
\partial^\sigma h^{\alpha\beta} - \partial^\sigma h^{\lambda\rho}
\,h^{\alpha\beta}\right).
\end{equation}

\subsection{Triangle diagram: odd parity}
\label{ssec:3ptTtriangle}

It is constructed with three $V_{gff}$ vertices. It has again an incoming line
with
momentum $q=k_1+k_2$ with Lorentz indices $\mu,\nu$. The two outgoing lines have
momenta $k_1,k_2$ with Lorentz labels $\lambda,\rho$ and $\alpha,\beta$,
respectively. The contribution is formally written as
\begin{multline} \label{T1}
\tilde T^{(1)}_{\mu\nu\alpha\beta\lambda\rho}(k_1,k_2) =-\frac 1 {512}\int
\frac
{d^3p}{(2\pi)^3}\, {\rm tr} \left[\left(\frac
1{\slashed{p}-m}\bigl((2p-k_1)_\lambda
\gamma_\rho+(\lambda\leftrightarrow \rho)\bigr)\right.\right.\frac
1{\slashed{p}-\slashed{k}_1-m}\\
\times \bigl((2p-2 k_1 -
k_2)_{\alpha}\gamma_{\beta}+(\alpha\leftrightarrow \beta)\bigr)
\left.\left.\frac 1{\slashed{p} - \slashed{q}-m}\bigl( (2{p}
-{q})_\mu\gamma_\nu+(\mu\leftrightarrow \nu)\bigr)\right)  \right],
\end{multline}
to which the cross graph contribution $\tilde T^{(2)}_{\mu\nu\alpha\beta\lambda\rho}(k_1,k_2)=\tilde T^{(1)}_{\mu\nu\lambda\rho\alpha\beta}(k_2,k_1)$ has to be added.

The odd part of (\ref{T1}) is 
\begin{multline} \label{T1odd}
\tilde T^{(1,odd)}_{\mu\nu\alpha\beta\lambda\rho}(k_1,k_2) =-\frac m {512}\int
\frac{d^3p}{(2\pi)^3}\, {\rm tr} \left[\slashed{p} \gamma_\rho
(\slashed{p}-\slashed{k}_1)\gamma_\beta\gamma_\nu+ 
 \gamma_\rho(\slashed{p}-\slashed{k}_1)\gamma_\beta(\slashed{p} -
\slashed{q})\gamma_\nu\right.\\
+\left.\slashed{p}\gamma_\rho\gamma_\beta(\slashed{p} -
\slashed{q}\gamma_\nu+m^2 \gamma_\rho\gamma_\beta\gamma_\nu\right]
\frac {(2p-k_1)_\lambda (2p-2k_1-k_2)_\alpha
(2p-q)_\mu}{(p^2-m^2)((p-k_1)^2-m^2)((p-q)^2-m^2)},
\end{multline}
where the symmetrization $\lambda\leftrightarrow \rho, \alpha\leftrightarrow
\beta, \mu\leftrightarrow \nu$ is understood.
In order to work out (\ref{T1odd}) we introduce two Feynman parameters: $u$
integrated between 0 and 1, and $v$
integrated between 0 and $1-u$. The denominator in  (\ref{T1odd}) becomes
\begin{equation}
 \left[ (p-(1-u)k_1-vk_2)^2+u(1-u)k_1^2+v(1-v)k_2^2 +2uv\, k_1\!\cdot\!
k_2-m^2\right]^3\0.
\end{equation}
After taking the traces, we get
\begin{eqnarray} \label{T1odd2}
\tilde T^{(1,odd)}_{\mu\nu\alpha\beta\lambda\rho}(k_1,k_2) &=&\frac {im}
{128}\int_0^1 du\int_0^{1-u} dv \int \frac
{d^3p}{(2\pi)^3}\, \left[  \epsilon_{\rho\sigma\nu} \left(-2p_\beta
k_1^\sigma
+k_{1\beta} q^\sigma+ q_\beta k_1^\sigma\right)\right.\0\\
&&+\left. 2\epsilon_{\sigma\beta\nu} p_\rho k_2^\sigma +\epsilon_{\rho\beta\nu}
\left(-5p^2 +(2p-q)\!\cdot\! k_1 
+m^2\right)+ \eta_{\rho\nu} \epsilon_{\sigma\beta\tau} k_1^\sigma k_2^\tau
\right]\0\\
&&\cdot \frac {(2p-k_1)_\lambda (2p-2k_1-k_2)_\alpha
(2p-q)_\mu}{[(p-(1-u)k_1-vk_2)^2+\Delta)^3},
\end{eqnarray}
where $\Delta= u(1-u)\,k_1^2+v(1-v)\,k_2^2 +2uv\, k_1\!\cdot\! k_2-m^2$.

So we can shift $p\to p'=p-(1-u)k_1-vk_2$ and integrate over $p'$. 
The $p$-integrals can be easily carried out, see Appendix \ref{sec:usefulint}. 
Unfortunately we are not able to integrate over $u$ and $v$ in an elementary
way.
So, one way to proceed is to use Mathematica, which however is not able to
integrate over both $u$
and $v$ unless some simplification is assumed. Therefore we choose the condition
$k_1^2=0=k_2^2$.  In this case
the total energy of the process is $E=\sqrt{q^2}= \sqrt {2 k_1\!\cdot\! k_2}$.

An alternative way is to use Mellin-Barnes representation for the propagators in (\ref{T1odd}) and proceed in an analytic as suggested by Boos and Davydychev, \cite{BoosDavy}. This second approach is discussed in Appendix \ref{sec:Lauricella}. In all the cases we were able to compare the two methods they give the same results (up to trivial terms).

\subsection{The IR limit}

The IR limit corresponds to $m\to\infty$ or, better, $\frac m{E}
\to\infty $ where $E=\sqrt{2k_1\!\cdot\!k_2}$. In this limit we find one
divergent
term ${\cal O}(m^2)$ and a series in the parameter $\frac m{E}$
starting from the 0-th order term.
The  ${\cal O}(m^2)$ term is (after adding the cross contribution)
\begin{multline} \label{3p-o-div}  
\sim \frac {m^2}{16\pi}\Big{[}16 \epsilon_{\sigma\beta\nu} k_2^\sigma \left(
\eta_{\rho\lambda}\eta_{\alpha\mu}
+ \eta_{\rho\alpha}\eta_{\lambda\mu}+
\eta_{\rho\mu}\eta_{\lambda\alpha}\right) +16\epsilon_{\sigma\rho\nu} k_1^\sigma
\left( \eta_{\beta\lambda}\eta_{\alpha\mu}
+ \eta_{\beta\alpha}\eta_{\lambda\mu}+
\eta_{\beta\mu}\eta_{\lambda\alpha}\right) \\
\left.-\epsilon_{\rho\beta\nu}\left(-\frac {112}3 (k_1-k_2)_\mu
\eta_{\lambda\alpha} +\frac {16}3 (11k_1+7k_2)_\alpha \eta_{\lambda\mu}- 
\frac {16}3(7k_1+11k_2)_\lambda \eta_{\alpha\mu}\right)\right].
\end{multline}
This term has to be symmetrized under $\mu\leftrightarrow \nu, \lambda
\leftrightarrow \rho,\alpha 
\leftrightarrow \beta$. It is a (non-conserved) local term.  It must be
subtracted from the action. 
Once this is done the relevant term for us is the 0-th order one. Let us call it
$\tilde T^{(odd,IR)}_{\mu\nu\alpha\beta\lambda\rho}(k_1,k_2)$. Its lengthy
explicit form
is written down in Appendix \ref{sec:3rdorderCS2}. 

If we compare this expression plus the contribution from the bubble diagram
with the one obtained from $CS_g^{(3)}$ in
\ref{ssec:3rdorderCS1},
which it is expected to reproduce, we see that they are different. This is not
surprising
in view of the discussion of the gauge case: the next to leading order of the
relevant CS
action is not straightaway reproduced by the relevant 3-point correlators, but need
corrections.
This can be seen also by contracting  $\tilde
T^{(odd,IR)}_{\mu\nu\alpha\beta\lambda\rho}(k_1,k_2)$
with $q^\mu$ and inserting it in the WI (\ref{3ptgravityWI}): the latter is
violated.

Now let us Fourier antitransform $\tilde
T^{(odd,IR)}_{\mu\nu\alpha\beta\lambda\rho}(k_1,k_2)$
and insert the result in the $W[g]$ generating function. We obtain a local
action term of 3rd order
in $h_{\mu\nu}$, which we may call $\widetilde{CS}_g^{(3)}$. Having in mind
(\ref{pertgaugeinv}), we find instead
\be \label{wrongWI}
\delta_\xi^{(1)} CS_g^{(2)}+\delta_\xi^{(0)} \widetilde{CS}_g^{(3)}= {\cal
Y}^{(2)}(\xi)\neq 0,
\ee
where ${\cal Y}^{(2)}(\xi)$ is an integrated local expression quadratic in
$h_{\mu\nu}$ and linear 
in the diffeomorphism parameter $\xi$. It is clear that in order to reproduce
(\ref{pertgaugeinv}) we must add
counterterms to the action, as we have done in the analogous case in
section \ref{sec:3ptcurrent}.
The question is whether this is possible. We can proceed as follows, we subtract
from (\ref{wrongWI}) 
the second equation in  (\ref{3ptgravityWI}) and obtain
\be \label{Ycounterterm}
\delta_\xi^{(0)} (\widetilde{CS}_g^{(3)}-CS_g^{(3)})= {\cal Y}^{(2)}(\xi).
\ee
Therefore $\widetilde{CS}_g^{(3)}-CS_g^{(3)}$ is the counterterm we have to subtract
from the action in order to
satisfy (\ref{3ptgravityWI}) and simultaneously reconstruct the gravitational
Chern-Simons action up to the third order. 
The procedure seems to be tautological, but this is simply due to the fact that
we already know the covariant answer, 
i.e. the gravitational CS action, otherwise we would have to work our way
through a painful analysis 
of all the terms in $ {\cal Y}^{(2)}(\xi)$ and find the corresponding
counterterms\footnote{Of course in the process of defining
the regularization and the IR limiting procedure, we are allowed to subtract all
the necessary counterterms   
(with the right properties) except fully covariant action terms (like the CS
action itself, for one).}.

The just outlined procedure is successful but somewhat disappointing. For the purpose of
reproducing $ CS_g^{(3)}$ the overall three-point calculation seems to be rather ineffective. 
One can say that the final result is completely determined by the two-point function analysis. 
Needless to say it would be preferable to find a regularization as well as a way
to take the IR and UV limits that do not break covariance. 
We do not know it this is possible. 

On the other hand the three-point function analysis is important for other reasons. For brevity
we do not report other explicit formulas about the coefficients of the expansion in $\frac Em$ 
and $\frac mE$. They all look like correlators, which may be local and non-local. The analysis 
of these expressions opens a new subject of investigation.

\section{Conclusion}
\label{sec:conclusion}

In this paper we have calculated two- and three-point functions of currents in the
free massive fermion model in 3d. We have mostly done our calculations with two different methods, as explained in 
Appendix \ref{sec:usefulint} and \ref{sec:Lauricella}, and obtained the same results. When the model is 
coupled to an external gauge potential and metric, respectively,
we have extracted from them, in the UV and the IR limit, CS action terms  
for gauge and gravity in 3d. We have also coupled the massless fermion model to higher spin potentials and explicitly
worked out the spin 3 case, by obtaining 
a very significant new result in the UV limit: the action reconstructed from the two-point current correlator
is a particular case of higher spin action introduced a long time ago by \cite{pope}; this is one of the possible generalizations 
of the CS action to higher spin. It is of course expected that carrying out analogous calculations for higher spin 
currents we will obtain the analogous generalizations of Chern-Simons to higher spin. Our result for the spin 3 case in the IR
is an action with a higher spin gauge symmetry, different from the UV one; we could not recognize it as a well-known higher spin action. 

Beside the results concerning effective actions terms in the UV and IR limit, there are other interesting aspects
of the correlators we obtain as intermediate steps. For instance, the odd parity current correlators at fixed points 
are conformal invariant and are limits of a free theory, but they cannot be obtained from any free theory
using the Wick theorem. There are also other interesting and not understood aspects. For instance, the two-point e.m. tensor 
correlators of the massive model
can be expanded in series of $E/m$ or $m/E$, where $E$ is the relevant energy, the coefficients in the expansion
being proportional always to the same conformal correlator. For the three-point functions the situation is more complicated, 
there is the possibility
of different limits and the expansion coefficients are also nonlocal. Still, however, we have a stratification similar to the 
one in the two-point functions with coefficient that look like
conserved three-point correlators (but have to be more carefully evaluated). One would like to know 
what theories these correlators refer to.

Finally it would be interesting to  embed the massive fermion model in an $AdS_4$ geometry.   
One can naively imagine the $AdS_4$ space foliated by 3d submanifolds with constant geodesic distance from the boundary
and a copy of the theory defined on each submanifold with a mass depending of the distance. This mass could be generated, for instance,
by the vev of a pseudoscalar field. This and the previous question certainly deserve further investigation. 

\bigskip

\noindent
{\bf \large Acknowledgements}

\bigskip

\noindent
One of us (L.B.) would like to thank Dario Francia and Mikhail Vasiliev for useful discussions. The research has been supported by Croatian Science Foundation under the project No.~8946 and by University of Rijeka under the research support No.~13.12.1.4.05.

\vspace{15pt}

\appendix
\section*{Appendices}

\section{Gamma matrices in 3d}
\label{sec:gammamatrix}

In $2+1$ dimensions we may take the gamma matrices, \cite{Parker:2009}, as 
\begin{equation}
\gamma_{0}=\sigma_{2}=\left(\begin{array}{cc}
0 & -i\\
i & 0
\end{array}\right),\quad\gamma_{1}=-i\sigma_{1}=\left(\begin{array}{cc}
0 & -i\\
-i & 0
\end{array}\right),\quad\gamma_{2}= i\sigma_{3}=\left(\begin{array}{cc}
i & 0\\
0 & -i
\end{array}\right).\label{eq:GammaRep}
\end{equation}
They satisfy the Clifford algebra relation
for the anticommutator of gamma matrices, namely
\[
\left\{ \gamma_{\mu},\gamma_{\nu}\right\} =2\eta_{\mu\nu},
\]
For the trace of three gamma matrices we have
\[
\trace\left(\gamma_{\mu}\gamma_{\nu}\gamma_{\rho}\right)=-2i\epsilon_{\mu\nu\rho
},
\]

\paragraph{Properties of gamma matrices in 3d}

\begin{gather*}
\trace\left(\gamma_{\mu}\gamma_{\nu}\right)=2\eta_{\mu\nu},\\
\trace\left(\gamma_{\mu}\gamma_{\nu}\gamma_{\rho}\right)=-2i\epsilon_{\mu\nu\rho
},\\
\trace\left(\gamma_{\mu}\gamma_{\nu}\gamma_{\rho}\gamma_{\sigma}
\right)=2\left(\eta_{\mu\nu}\eta_{\rho\sigma}-\eta_{\mu\rho}\eta_{\nu\sigma}
+\eta_{\mu\sigma}\eta_{\nu\rho}\right),\\
\left\{ \gamma^{\mu},\Sigma^{\rho\sigma}\right\} =-i\epsilon^{\mu\rho\sigma}.
\end{gather*}

\begin{equation}
 \gamma_\sigma \gamma_\mu\gamma_\nu= -i\epsilon_{\sigma\mu\nu} +\eta_{\mu\sigma}
\gamma_\nu 
-\eta_{\sigma\nu} \gamma_\mu + \eta_{\mu\nu} \gamma_\sigma 
\end{equation}

\begin{equation}
\trace\left(\gamma_\sigma \gamma_\mu\gamma_\nu\gamma_\lambda \gamma_\rho\right)
= -2i\left( \epsilon_{\mu\nu\lambda} \eta_{\sigma\rho}
+\eta_{\mu\nu} \epsilon_{\sigma\lambda\rho} -\eta_{\mu\lambda}
\epsilon_{\sigma\nu\rho}  + \eta_{\nu\lambda} \epsilon_{\sigma\mu\rho}\right)
\end{equation}

\paragraph{Identity for $\epsilon$ and $\eta$ tensors}:
\begin{equation}
 \eta_{\mu\nu}  \epsilon_{\lambda\rho\sigma}-\eta_{\mu\lambda} \epsilon_{
\nu\rho\sigma}+
\eta_{\mu\rho} \epsilon_{\nu\lambda\sigma}-\eta_{\mu\sigma}
\epsilon_{\nu\lambda\rho} =0\nonumber
\end{equation}

Finally, to make contact with the spinorial label notation of ref.\cite{pope}
one may use 
the symmetric matrices
\be 
(\tilde \gamma_0)_{\alpha\beta} = i (\gamma_0)_\alpha{}^\gamma
\epsilon_{\gamma\beta},\quad\quad
(\tilde \gamma_1)_{\alpha\beta} = (\gamma_1)_\alpha{}^\gamma
\epsilon_{\gamma\beta},\quad\quad
(\tilde \gamma_2)_{\alpha\beta} =- (\gamma_2)_\alpha{}^\gamma
\epsilon_{\gamma\beta},\label{gammatiilde}
\ee
where $\epsilon$ is the antisymmetric matrix with $\epsilon_{12}=-1$, and write
\be 
h_{\alpha_1\alpha_2\alpha_3\alpha_4\alpha_5\alpha_6} = h_{abc} (\tilde
\gamma^a)_{\alpha_1\alpha_2} (\tilde \gamma^b)_{\alpha_3\alpha_4}
 (\tilde \gamma^c)_{\alpha_5\alpha_6}, \quad\quad \partial_a (\tilde
\gamma^a\epsilon)_{\alpha}{}^{\beta} = \partial_\alpha{}^\beta,\quad\quad
{\rm etc.}
\ee

\paragraph{Wick rotation}

Among the various conventions for the Wick rotation to compute Feynman diagram,
we think the simplest one is 
given by the following formal rule on the metric: $\eta_{\mu\nu}\to
-\eta^{(E)}_{\mu\nu}$. This implies
\be 
k^2\to -(k^{(E)})^2, \quad\quad p_\mu p_\nu \to \frac 13
(p^{(E)})^2\,\eta^{(E)}_{\mu\nu},\ldots\0
\ee
We have also to multiply any momentum integral by $i$. For the sake of
simplicity we always understand the superscript $^{(E)}$.

\section{Invariances of the 3d free massive fermion}
\label{sec:invariances}

In the theory defined by \ref{actiong} there is a problem connected
with  the presence of $\sqrt g=e$ in the action. When defining 
the Feynman rules we face two possibilities: 1) either we incorporate $\sqrt e$
in the spinor field $\psi$, so that the factor $\sqrt g$ in 
fact disappears from the action, or, 2), we keep the action as it is.

In the first case we define a new field $\Psi= \sqrt{e}\psi$. The new action
becomes
\begin{equation}
S=\int
d^{3}x\,\left[i\bar{\Psi}E_{a}^{\mu}\gamma^{a}\nabla_{\mu}\Psi-m\bar{\Psi}
\psi\right].\label{actionPsi}
\end{equation}
due essentially to the fact that $\nabla_\lambda g_{\mu\nu}=0$. The action is
still diff-invariant provided $\Psi$ transforms as
\begin{equation}
\delta_\xi \Psi= \xi^\mu\partial_\mu \Psi +\frac 12 \nabla_\mu \xi^\mu
\Psi\label{Psidiff}
\end{equation}
In the case $m=0$ we also have Weyl invariance with
\begin{equation}
 \delta_\omega \Psi=\frac 12 \omega \Psi, \quad\quad {\rm instead\,\,\,
of}\quad\quad   \delta_\omega \psi= \omega \psi,\label{PsiWeyl}
\end{equation}
So the simmetries are classically preserved while passing from $\psi$ to $\Psi$.
From a quantum point of view this might seem a Weyl transformation of $\Psi$,
but
it is not accompanied by a corresponding Weyl transformation of the metric. So it
is simply a field redefinition, not a symmetry operation.
 
Alternative 1) is the procedure of Delbourgo-Salam. The action can be rewritten
\begin{equation}
S=\int
d^{3}x\,\left[\frac{i}{2}\bar{\Psi}E_{a}^{\mu}\gamma^{a}\stackrel{
\leftrightarrow}{\partial}\Psi-m\bar{\Psi}\Psi
+\frac{1}{2}E_{a}^{\mu}\omega_{\mu
bc}\epsilon^{abc}\bar{\Psi}\Psi\right].\label{actionPsi1}
\end{equation}
In this case we have one single graviton-fermion-fermion vertex $V_{gff}$
represented by
\begin{equation}
 \frac i8 \left[(p+p')_\mu \gamma_\nu +(p+p')_\nu \gamma_\mu\right]\label{Vgff}
\end{equation}
and one single 2-gravitons--2-fermions vertex $V_{ggff}$ given by
\begin{equation}
 \frac 1{16} t_{\mu\nu\mu'\nu'\lambda} (k-k')^\lambda \label{Vggff}
\end{equation}
where
\begin{equation}
t_{\mu\nu\mu'\nu'\lambda}=\eta_{\mu\mu'}\epsilon_{\nu\nu'\lambda}
+\eta_{\nu\mu'}\epsilon_{\mu\nu'\lambda} +
\eta_{\mu\nu'}\epsilon_{\nu\mu'\lambda}+
\eta_{\nu\nu'}\epsilon_{\mu\mu'\lambda}, \label{t}
\end{equation}
the fermion propagator being
\begin{equation}
\frac i{\slashed{p} -m+i\epsilon}\nonumber
\end{equation}
The convention for momenta are the same as in \cite{BGL,BDL}.

Alternative 2) introduces new vertices. In this case the Lagrangian can be
written
\begin{eqnarray}
L&=&\frac i2 \bar \psi\, \gamma^a \stackrel{\leftrightarrow}{\partial}_a \psi
-i\,m\, \bar\psi \psi\nonumber\\
&& + \frac i4    \bar \psi \,\gamma^a h_a^\mu
\stackrel{\leftrightarrow}{\partial}_\mu \psi+
\frac i4 h_\lambda^\lambda\, \bar \psi\, \gamma^a
\stackrel{\leftrightarrow}{\partial}_a \psi 
-\frac i2  h_\lambda^\lambda \,m\, \bar\psi \psi \label{Lag}\\
&& + \frac i8  h_\lambda^\lambda \, \bar \psi\, \gamma^a\, h_a^\mu
\stackrel{\leftrightarrow}{\partial}_\mu \psi -
\frac 1{16} \bar \psi\, h_c^\lambda \partial_a h_{\lambda b} \psi\,
\epsilon^{abc}\nonumber
\end{eqnarray}
As a consequence we have three new vertices.  A vertex $V'_{gff}$ coming from
the mass term
\begin{equation}
 -\frac i2 \,m\, \eta_{\mu\nu} {\bf 1},\label{Vgff'}
\end{equation}
another $V''_{gff}$ coming from the kinetic term
\begin{equation}
 \frac i4 \eta_{\mu\nu} (\slashed{p} + \slashed{p}')\label{Vgff''}
\end{equation}
and a new $V'_{ggff}$
\begin{equation}
 \frac i8 \eta_{\mu'\nu'} \left[(p+p')_\mu \gamma_\nu +(p+p')_\nu
\gamma_\mu\right]\label{Vggff'}
\end{equation}

An obvious conjecture is that the two procedures lead to the same results, up to
trivial terms. But this has still to be proved. 

In this paper we follow alternative 1 only.

\section{Perturbative cohomology}
\label{sec:pertcohomology}

In this Appendix we define the form of local cohomology which is needed in
perturbative field theory. Let us start from the gauge transformations.
\be
\delta A = d\lambda + [A,\lambda], \quad\quad \delta \lambda =-\frac 12
[\lambda,\lambda]_+,\quad\quad \delta^2=0,\quad\quad 
\lambda = \lambda^a(x) T^a \label{gaugetransf}
\ee
To dovetail the perturbative expansion it is useful to split it. Take $A$ and
$\lambda$ infinitesimal and define the perturbative cohomology
\be 
&&\delta^{(0)} A = d\lambda, \quad\quad \delta^{(0)}\lambda=0, \quad\quad 
(\delta^{(0)})^2=0\0\\
&&\delta^{(1)} A = [A,\lambda],  \quad\quad  \delta^{(1)}\lambda=-\frac 12
[\lambda,\lambda]_+\0\\
&& \delta^{(0)} \delta^{(1)}+\delta^{(1)} \delta^{(0)}=0,\quad\quad
(\delta^{(1)})^2=0\label{pertcoho}
\ee

The full coboundary operator for diffeomorphisms is given by the transformations
\begin{equation}
\delta_\xi g_{\mu\nu}=\nabla_\mu \xi_\nu + \nabla_\nu \xi_\nu, \quad\quad 
\delta_\xi \xi^\mu = \xi^\lambda \partial_\lambda \xi^\mu\label{fullcoho}
\end{equation}
with $\xi_\mu = g_{\mu\nu}\xi^\nu$.
We can introduce a perturbative cohomology, or graded cohomology, using as
grading the order of infinitesimal, as follows
\begin{eqnarray}
g_{\mu\nu} = \eta_{\mu\nu} + h_{\mu\nu}, \quad\quad g^{\mu\nu}= \eta^{\mu\nu}
-h^{\mu\nu}+ h^\mu_\lambda h^{\lambda\nu}+\ldots \label{metricapprox}
\end{eqnarray}
The analogous expansions for the vielbein is 
\begin{equation}
 e_\mu^a = \delta_\mu^a+\chi_\mu^a+\frac 12 \psi_\mu^a+\ldots,  \0
\end{equation}
Since $e_\mu^a \eta_{ab}e^b_\nu=h_{\mu\nu}$, we have
\begin{equation}
\chi_{\mu\nu}= \frac 12 h_{\mu\nu},\quad\quad \psi_{\mu\nu} = 
-\chi_\mu^a\chi_{a\nu}= -\frac 14 h_\mu^\lambda h_{\lambda\nu},\quad\quad \ldots
\label{dreibein}
\end{equation}
This leads to the following expansion for the spin connection
$\omega_{\mu}^{ab}$
\begin{eqnarray}
 \omega_{\mu}^{ab}&=& \frac 12 e^{\nu a} \left(\partial_\mu e_\nu^b -
\partial_\nu e_\mu^b\right) 
-   \frac 12 e^{\nu b} \left(\partial_\mu e_\nu^a - \partial_\nu e_\mu^a\right)-
\frac 12 e^{\rho a}  e^{\sigma b} \left(\partial_\rho e_{\sigma c} -
\partial_\sigma e_{\rho c}\right)e^c_\mu\label{spiconn}\\
&=& -\frac 12 \left(\partial^a h_\mu^b-\partial^b h_\mu^a\right) -
\frac 18  \left(h^{\sigma a}\partial_\mu h_\sigma^b-h^{\sigma b}\partial_\mu
h_\sigma^a\right) +
\frac 14 \left(h^{\sigma a}\partial_\sigma h_{\mu}^b -h^{\sigma
b}\partial_\sigma h_{\mu}^a\right) \0\\
&& - \frac 18  \left(h^{\sigma a}\partial_\sigma h_{\mu}^b -h^{\sigma
b}\partial_\sigma h_\sigma^a\right) -
\frac 18 h_\mu^c\left(\partial ^a h^b_c -\partial ^b h^a_c\right) \0\\
&&-\frac 18 \left(\partial^b (h_\mu^\lambda h_\lambda ^a) - \partial^a
(h_\mu^\lambda h_\lambda^b)\right)+\ldots\0
\end{eqnarray}

Inserting the above expansions in (\ref{fullcoho}) we see that we have a grading
in
the transformations, given by the order of infinitesimals.
So we can define a  sequence of transformations
\begin{equation}
\delta_\xi= \delta^{(0)}_\xi+\delta^{(1)}_\xi+\delta^{(2)}_\xi+\ldots \nonumber
\end{equation}
At the lowest level we find immediately
\begin{eqnarray}
&& \delta^{(0)}_\xi h_{\mu\nu}= \partial_\mu\xi_\nu + \partial_\nu
\xi_\mu,\quad\quad  \delta^{(0)}_\xi \xi_\mu=0\label{0cohomo}
\end{eqnarray}
and $\xi_\mu = \xi^\mu$. Since $(\delta^{(0)}_\xi)^2=0$ this defines a
cohomology problem. 

At the next level we get
\begin{equation}
\delta_\xi^{(1)} h_{\mu\nu}   =  \xi^\lambda
\partial_\lambda h_{\mu\nu} +
\partial_\mu \xi^\lambda h_{\lambda\nu} + 
\partial_\nu \xi^\lambda h_{\mu\lambda}, \quad \quad \delta^{(1)}_\xi \xi^\mu =
\xi^\lambda \partial_\lambda \xi^\mu\label{1cohomo}
\end{equation}
 
One can verify that
\begin{equation}
 (\delta^{(0)}_\xi)^2=0\quad\quad   \delta^{(0)}_\xi\delta^{(1)}_\xi
+\delta^{(1)}_\xi\delta^{(0)}_\xi =0,\quad\quad
(\delta^{(1)}_\xi)^2=0\label{xinilpotency}
\end{equation}

Proceeding in the same way we can define an analogous sequence of
transformations for the Weyl transformations. 
From $g_{\mu\nu} = \eta_{\mu\nu} + h_{\mu\nu} $
and $\delta_\omega h_{\mu\nu}= 2\omega g_{\mu\nu}$ we find
\begin{eqnarray}
&&\delta^{(0)}_\omega h_{\mu\nu} = 2\omega \eta_{\mu\nu},\quad\quad
\delta^{(1)}_\omega k_{\mu\nu}= 2\omega h_{\mu\nu},\quad\quad
\delta^{(2)}_\omega h_{\mu\nu}= 0,\ldots \label{deltaomegah}
\end{eqnarray}
as well as $ \delta^{(0)}_\omega \omega=\delta^{(1)}_\omega \omega=0,...$.

Notice that we have 
$\delta^{(0)}_\xi \omega=0, \delta^{(1)}_\xi \omega=\xi^\lambda \partial_\lambda
\omega$.
As a consequence we can extend (\ref{xinilpotency}) to
\begin{equation}
(\delta^{(0)}_\xi +\delta^{(0)}_\omega) (\delta^{(1)}_\xi +\delta^{(1)}_\omega)+
(\delta^{(1)}_\xi +\delta^{(1)}_\omega) (\delta^{(0)}_\xi
+\delta^{(0)}_\omega)=0\label{couplednilpotency}
\end{equation}
and $\delta^{(1)}_\xi\delta^{(1)}_\omega+\delta^{(1)}_\omega
\delta^{(1)}_\xi=0$,
which together with the previous relations make
\begin{equation}
(\delta^{(0)}_\xi +\delta^{(0)}_\omega+\delta^{(1)}_\xi
+\delta^{(1)}_\omega)^2=0 \label{fullcouplednilpotency}
\end{equation}

For what concerns the higher tensor field $B_{\mu\nu\lambda}$ in this paper we
use only the lowest order transformations given by
(\ref{Bgaugetransf}) and (\ref{Bweyltransf}).

\section{Useful integrals}
\label{sec:usefulint}

The Euclidean integrals over the momentum $p$ we use for the 2-point function are:
\begin{eqnarray}
\int\frac{d^{3}p}{\left(2\pi\right)^{3}}\frac{1}{\left(p^{2}
+\Delta\right)^{2}} & = & \frac{1}{8\pi}\frac{1}{\sqrt{\Delta}},\label{Eintegrals1}\\
\int\frac{d^{3}p}{\left(2\pi\right)^{3}}\frac{p^{2}}{\left(p^{2}
+\Delta\right)^{2}} & = & -\frac{3}{8\pi}\sqrt{\Delta},\label{Eintegrals2}\\
\int\frac{d^{3}p}{\left(2\pi\right)^{3}}\frac{p^{4}}{\left(p^{2}
+\Delta\right)^{2}} & = & \frac{5}{8\pi}\Delta^{3/2},\label{Eintegrals3}
\end{eqnarray} 
where
$\Delta=m^{2}+x\left(1-x\right)k^{2} $ 
and for the 3-point functions
\begin{eqnarray}
\int \frac {d^3p}{(2\pi)^3} \frac 1{(p^2+\Delta)^3} &=& \frac 1{32\pi} \frac
1{\Delta^{\frac 32}}\\
\int \frac {d^3p}{(2\pi)^3} \frac {p^2}{(p^2+\Delta)^3} &=& \frac 3{32\pi} \frac
1{\sqrt \Delta}\\
\int \frac {d^3p}{(2\pi)^3} \frac {p^4}{(p^2+\Delta)^3} &=& \frac {15}{32\pi} 
\sqrt{\Delta}
\end{eqnarray} 
where $\Delta=m^{2}+u(1-u)k_1^2+v(1-v)k_2^2 +2uv k_1\!\cdot\! k_2$.  In these
formulae $x,u,v$
are Feynman parameters.

\paragraph{Sample calculation}

As an example of our calculations we explain here some details of the derivation
in \ref{ssec:TT}. To make sense of the integral in (\ref{Tmnlrodd}) we have to
go Euclidean, which
implies
$p^2\to -p^2, k^2\to -k^2, \eta_{\mu\nu}\to -\eta_{\mu\nu}$ and $d^3p \to i
d^3p$. Therefore
\begin{equation} \label{Tmnlrodd1}
\tilde T^{(odd)}_{\mu\nu\lambda\rho}(k) = \frac m{32} \int_0^1dx\, \int\frac
{d^3p}{(2\pi)^3}\left[ \epsilon_{\sigma \nu\rho}\, k^\sigma  \, \frac { \frac
43 p^2\eta_{\mu\lambda} +(2x-1)^2k_\mu k_\lambda }{[p^2+m^2 +x(1-x)k^2]^2}+
\left(\begin{matrix}{\mu\leftrightarrow \nu}\\{\lambda\leftrightarrow
\rho}\end{matrix}\right) \right].
\end{equation}
Next we use the appropriate Euclidean integrals above to integrate
over $p$ and  get 
\begin{multline} \label{Tmnlrodd2}
\tilde T^{(odd)}_{\mu\nu\lambda\rho}(k) = -\frac m{256\pi}
\int_0^1dx\, \epsilon_{\sigma \nu\rho}\, k^\sigma  \\ \times  \left(4
\eta_{\mu\lambda}(m^2+x(1-x) k^2)^{\frac 12} +k_\mu k_\lambda
\frac{(2x-1)^2}{(m^2+x(1-x) k^2)^{\frac 12}}\right) 
+\left(\begin{matrix}{\mu\leftrightarrow \nu}\\{\lambda\leftrightarrow \rho}\end{matrix}\right) 
\end{multline}
The $x$ integrals are well defined:
\begin{eqnarray}
\int_0^1dx \, (m^2+x(1-x) k^2)^{\frac 12}&=& \frac 12 m +\frac 14 \frac
{k^2+4m^2}{|k|} \arctan\frac{|k|}{2m}\label{xint1}\\
\int_0^1dx \, \frac{(2x-1)^2}{(m^2+x(1-x) k^2)^{\frac 12}}  &=& -2\frac m{k^2}
+\frac {k^2+4m^2}{|k|^3} \arctan\frac{|k|}{2m}\label{xint2}
\end{eqnarray}
Therefore the result is
\begin{eqnarray}
\tilde T^{(odd)}_{\mu\nu\lambda\rho}(k) &=& \frac m{256\pi}\, \epsilon_{\sigma
\nu\rho}\, k^\sigma  \,\left[ -\eta_{\mu\lambda}\left( 2m+  \frac
{k^2+4m^2}{|k|}
\arctan \frac{|k|}{2m}\right)\right.\nonumber\\
&& +\left. \frac {k_\mu k\nu}{k^2} \left(- 2m+  \frac {k^2+4m^2}{|k|} \arctan
\frac{|k|}{2m}\right)\right]+ \left(\begin{matrix}{\mu\leftrightarrow
\nu}\\{\lambda\leftrightarrow \rho}\end{matrix}\right)\0 \\
&=&\frac m{256\pi}\, \epsilon_{\sigma \nu\rho}\, k^\sigma 
\,\left[2m\left(-\eta_{\mu\lambda}-\frac {k_\mu
k_\lambda}{k^2}\right)\right.\0\\
&&\left. +
\left(-\eta_{\mu\lambda}+\frac {k_\mu k_\lambda}{k^2}\right)\frac
{k^2+4m^2}{|k|}
\arctan \frac{|k|}{2m}\right]+ \left(\begin{matrix}{\mu\leftrightarrow
\nu}\\{\lambda\leftrightarrow \rho}\end{matrix}\right) \label{Tmnlrodd3}
\end{eqnarray}
The final step is to return to the Lorentzian metric, $k^2\to -k^2$ and
$\eta_{\mu\nu}\to -\eta_{\mu\nu}$, $\arctan\frac{|k|}{2m} \to i\,\arctanh\frac{|k|}{2m}$.

\section{An alternative method for Feynman integrals}
\label{sec:Lauricella}

An alternative method to calculate Feynman diagrams was introduced in a series
of paper by A. I .Davydychev and collaborators, \cite{BoosDavy}. The basic
integral to be computed in our case are
\begin{equation}\label{J2initial}
J_2(d;\alpha,\beta;m)=\int\frac{d^d
p}{(2\pi)^d}\frac{1}{\left(p^2-m^2\right)^\alpha\left( (p-k)^2-m^2
\right)^\beta}
\end{equation}
and
\be\label{J3initial}
J_3(d;\alpha,\beta,\gamma;m)=\int\frac {d^d p}{(2\pi)^d}  \frac
1{(p^2-m^2)^\alpha((p-k_1)^2-m^2)^\beta ((p-q)^2-m^2)^\gamma},
\ee
with $q=k_1+k_2$. Following \cite{BoosDavy} these can be expressed, via the
Mellin-Barnes representation of the propagator, as
\begin{multline}\label{J2Mellin}
J_2(d;\alpha,\beta;m)=\frac{i^{1-d}}{(4\pi)^\frac{d}{2}}\frac{(-m^2)^{\frac{d}{2
}-\alpha-\beta}}{\Gamma\left( \alpha \right)\Gamma\left( \beta \right)} \\ 
\times\int \frac{du}{2\pi i}\left(-\frac{k^2}{m^2}\right)^u
\gammaFct{-u}\frac{\gammaFct{\alpha+u}\gammaFct{\beta+u}\gammaFct{
\alpha+\beta-\frac d2+u}}{\gammaFct{\alpha+\beta+2u}}
\end{multline}
and
\begin{multline}\label{J3Mellin}
J_3(d;\alpha,\beta,\gamma;m)= 
\frac {i^{1-d}}{(4\pi)^{\frac d2}}\frac{ (-m^2)^{\frac
d2-\alpha-\beta-\gamma}}{\Gamma(\alpha)\Gamma(\beta)\Gamma(\gamma)}\\
\times  \int \frac{ds}{2\pi i}\frac{dt}{2\pi i}\frac{du}{2\pi i}
\left(-\frac{k_1^2}{m^2}\right)^s\left(-\frac{q^2}{m^2}\right)^t\left(-\frac{
k_2^2}{m^2}\right)^u
\Gamma(-s) \Gamma(-t)\Gamma(-u)\\
\times \frac {\Gamma(\alpha+\beta+\gamma-\frac d2
+s+t+u)\Gamma(\alpha+s+t)\Gamma(\beta+s+u)
+\Gamma(\gamma+t+u)}{\Gamma(\alpha+\beta+\gamma  +2s+2t+2u)}.
\end{multline}
The integrals run from $-i\infty$ to  $i\infty$ along vertical contours that
separate the positive poles of the $\Gamma$'s from the negative ones. 
Positive poles are those of $\gammaFct{-u}$ in the case of $J_2$ or those of
$\Gamma(-s) \Gamma(-t)\Gamma(-u)$ in the case of $J_3$, negative poles are the
others.  It is clear that the contours of integration 
must cross the real axis just to the left of the origin. The contours close
either to the left or to the right
in such a way as to assure convergence of the series. Let us analyse more
closely the case of $J_2$ to better understand how this works. Using the
duplication formula of the gamma function, i.e.
\begin{equation}\label{dublingFormula}
\gammaFct{2z}=2^{2z-1}\pi^{-\frac{1}{2}}\gammaFct{z}\gammaFct{z+\frac 12}
\end{equation}
we are able to recast (\ref{J2Mellin}) into the form
\begin{multline}
\frac{i^{1-d}}{(4\pi)^\frac d2}(-m^2)^{\frac d2
-\alpha-\beta}\frac{\gammaFct{\frac{\alpha+\beta}{2}}\gammaFct{\frac{
\alpha+\beta+1}{2}}}{\gammaFct{\alpha}\gammaFct{\beta}\gammaFct{\alpha+\beta}}
\int \frac{du}{2\pi i}\lp -\frac{k^2}{4m^2} \rp^u\\
\times
\gammaFct{-u}\frac{\gammaFct{\alpha+u}\gammaFct{\beta+u}\gammaFct{
\alpha+\beta-\frac
d2+u}}{\gammaFct{\frac{\alpha+\beta}{2}+u}\gammaFct{\frac{\alpha+\beta+1}{2}+u}}
.
\end{multline}
Assuming $\left| \frac{k^2}{4m^2} \right|<1$ (IR region), we must close the
contour of integration on the right ($\mathrm{Re}\lp u\rp>0$) in order to
guarantee convergence of the result and by doing so we will pick-up the poles of
$\gammaFct{-u}$. For $\alpha=\beta=1$ and $d=3$ we obtain
\begin{equation}
J_2^{\rm IR}(3;1,1;m) = \frac{i}{8\pi
|m|}\sum_{j=0}^\infty\lp\frac{k^2}{4m^2}\rp^j\frac{1}{2j+1} = \frac{i}{4\pi |k|}
\mathrm{arctanh}\lp\sqrt{\frac{k^2}{4m^2}}\rp.
\end{equation}
On the other hand, assuming $\left| \frac{k^2}{4m^2} \right|>1$ (UV region), we
need to close the integration contour on the left. For $\alpha=\beta=1$ and
$d=3$, we will have poles at $u=-\frac 12$ and at $u=-1,-2,-3,\dots$, hence
\begin{align}
J_2^{\rm UV}(3;1,1;m) &= \frac{i}{8\pi|m|}\lp i\pi\frac{|m|}{|k|} +
\sum_{j=1}^\infty\lp \frac{4m^2}{k^2} \rp^j \frac{1}{(2j-1)} \rp \0 \\ 
&= -\frac{1}{8|k|}+\frac{i}{4\pi|k|}{\rm arctanh}\lp\sqrt{\frac{4m^2}{k^2}} \rp.
\end{align}

As far as (\ref{J3Mellin}) is concerned,
in this paper we are interested in particular in the IR region, which is the one
where $m^2$ is much larger than $k_1^2,k_2^2,q^2$ in the case of $J_3$. This
requires that the relevant powers $s,t,u$ in the integrands be positive, and,
so, the contours must close around the poles of the  positive real axis, that is
the poles of  $\Gamma(-s) \Gamma(-t)\Gamma(-u)$. An easy calculation gives
\begin{multline}
J_3(d;\alpha,\beta,\gamma;m)= \frac {i^{1-d}}{(4\pi)^{\frac d2}}  (-m^2)^{\frac
d2-\alpha-\beta-\gamma}
\frac {\Gamma(\alpha+\beta+\gamma-\frac
n2)}{\Gamma(\alpha+\beta+\gamma)}\label{J3IR}\\
\times \Phi_3\left[ \begin{matrix} \alpha+\beta+\gamma-\frac
n2,\alpha,\beta,\gamma\\
                 \quad\alpha+\beta+\gamma\\
                \end{matrix}\Big{\vert}
\frac{k_1^2}{m^2},\frac{q^2}{m^2},\frac{k_2^2}{m^2}\right],
\end{multline}
where $\Phi_3$ is a generalized Lauricella function:
\begin{multline}
 \Phi_3\left[ \begin{matrix}a_1 ,a_2,a_3,a_4\\
                 \quad c\\
                \end{matrix}\Big{\vert} z_1,z_2,z_3\right]\\
= \sum_{j_1=0}^\infty    \sum_{j_2=0}^\infty \sum_{j_2=0}^\infty
\frac {z_1^{j_1}}{j_1!} \frac {z_2^{j_2}}{j_2!} \frac {z_3^{j_3}}{j_3!}
\frac{(a_1)_{j_1+j_2+j_3} (a_2)_{j_1+j_2}  (a_3)_{j_1+j_3} (a_4)_{j_2+j_3}}
{(c)_{2j_1+2j_2+2j_3}}\label{Phi3}
\end{multline}
where $(a)_n=\frac {\Gamma(a+n)}{\Gamma(a)}$ is the Pochammer symbol. The
leading term in the IR is clearly
the one given by $j_1=j_2=j_3=0$, i.e. by setting $\Phi_3=1$ in (\ref{J3IR}).

In general, we need to evaluate not only (\ref{J3initial}) but more general
integrals 
\be
J_{3,{\mu_1}\dots {\mu_M} }(d;\alpha,\beta,\gamma;m)=\! \int \!\! \frac {d^d
p}{(2\pi)^d}  
\frac {p_{\mu_1}\dots p_{\mu_M} }{(p^2-m^2)^\alpha((p-k_1)^2-m^2)^\beta
((p-q)^2-m^2)^\gamma}\label{J3tensor}
\ee
One can prove by induction that the following formula holds in general
\begin{multline}
J_{3,\mu_{1}\dots\mu_{M}} \left(d; \alpha,\beta,\gamma;m \right) =
\sum_{\substack{\lambda,\kappa_{1},\kappa_2,\kappa_{3}\\2\lambda+\sum\kappa_{i}
=M
}
}\left(-\frac{1}{2}\right)^{\lambda}\left(4\pi\right)^{M-\lambda}\left\{
\left[\eta\right]^{\lambda}\left[q_{1}\right]^{\kappa_{1}} 
\left[q_{2}\right]^{\kappa_{2}}\left[q_{3}\right]^{\kappa_{3}}\right\}
_{\mu_{1}\dots\mu_{M}}\\
\times\pochhammer{\alpha}{\kappa_{1}}\pochhammer{\beta}{\kappa_{2}}\pochhammer{
\gamma}{\kappa_{3}}
J_3(d+2(M-\lambda);\alpha+\kappa_1,\beta+\kappa_2,\gamma+\kappa_3;m)
,\label{eq:GeneralTensorIntegralApdx}
\end{multline}
where the symbol $\left\{
\left[\eta\right]^{\lambda}\left[q_{1}\right]^{\kappa_{1}}\dots\left[q_{N}\right
]^{\kappa_{N}}\right\} _{\mu_{1}\dots\mu_{M}}$
stands for the complete symmetrization of the objects inside the curly brackts,
for example
\[
\left\{ \eta q_{1}\right\}
_{\mu_{1}\mu_{2}\mu_{3}}=\eta_{\mu_{1}\mu_{2}}q_{1\mu_{3}}+\eta_{\mu_{1}\mu_{3}}
q_{1\mu_{2}}+\eta_{\mu_{2}\mu_{3}}q_{1\mu_{1}}.
\]

\section{Third order gravity CS and 3-point e.m. correlator}
\label{sec:3rdorderCS}

In this appendix we collect the result concerning the odd parity 3-point function of
the e.m. tensor and 
its relation to the third order term in gravitational CS action.

\subsection{The third order gravitational CS}
\label{ssec:3rdorderCS1}

From the action term (\ref{CS3h}), by differentiating three times with respect
to
$h_{\mu\nu}(x)$,$h_{\lambda\rho}(y)$ and $h_{\alpha\beta}(z)$ and 
Fourier-transforming the result one gets the sum of the following local terms in
momentum space (they
feature in the same order they appear in (\ref{CS3h}), 
\begin{eqnarray}
 \frac{\kappa}{4}\,\frac i4\,k_1^\sigma k_2^\tau \!&&\!\Big{(} \epsilon_{\mu\sigma\tau}  (q_\alpha
\eta_{\nu\lambda}\eta_{\rho\beta} 
-q_\rho \eta_{\nu\alpha}\eta_{\lambda\beta}) + \epsilon_{\lambda\sigma\tau}
(k_{1\alpha} \eta_{\mu\rho}\eta_{\nu\beta}- k_{1\nu}  
\eta_{\mu\alpha}\eta_{\rho\beta} )\nonumber\\
&& +\epsilon_{\alpha\sigma\tau} (k_{2\nu}\eta_{\mu\rho}\eta_{\lambda\beta}
-k_{2\lambda}\eta_{\mu\beta}\eta_{\nu\rho} )\Big{)}\label{CS1}
\end{eqnarray}
\begin{eqnarray}
\frac{\kappa}{4}\,\frac i4 \epsilon_{\mu\lambda\alpha}\,\Big{(}
&& -  k_1\!\cdot \!k_2 \left( k_{1\rho}
\eta_{\beta\nu} - k_{2\beta} \eta_{\rho\nu} 
+(k_2-k_1)_\nu \eta_{\beta\rho}  \right) \0\\
&& 
+ k_2^2 \left(
  \eta_{\beta\rho} k_{1\nu}
- \eta_{\nu\rho} k_{1\beta}   \right)
+  k_1^2 \left( 
  \eta_{\beta\nu} k_{2\rho}
- \eta_{\beta\rho} k_{2\nu} \right)
\Big{)}
\label{CS2}
\end{eqnarray}
\begin{eqnarray}
\frac{\kappa}{4}\, \frac i4  \epsilon_{\mu\lambda\alpha} \left( k_{1\beta} q_\rho k_{2\nu}
-k_{1\nu} q_\beta k_{2\rho} \right) \label{CS3}
\end{eqnarray}
\begin{eqnarray}
\frac{\kappa}{4}\, \frac i4\,\Big{(}\!&&\!  
\epsilon_{\mu\alpha\sigma} 
   q^\sigma(q_\beta \eta_{\nu\lambda} -q_\lambda \eta_{\beta\nu}) k_{2\rho}+   
\epsilon_{\mu\lambda\sigma} 
   q^\sigma(q_\rho \eta_{\nu\beta}  -q_\beta \eta_{\nu\rho}) k_{1\alpha}\0\\
\!&&\!+  
\epsilon_{\lambda\alpha\sigma} 
  k_1^\sigma(k_{1\beta} \eta_{\mu\rho}-k_{1\mu} \eta_{\beta\rho}) k_{2\nu}+
\epsilon_{\alpha\lambda\sigma} 
  k_2^\sigma(k_{2\rho} \eta_{\mu\beta} -k_{2\mu} \eta_{\beta\rho}) k_{1\nu}\0\\
\!&&\!+
\epsilon_{\mu\lambda\sigma} 
  k_1^\sigma(k_{1\nu} \eta_{\beta\rho} -k_{1\beta} \eta_{\rho\nu}) q_\alpha+ 
\epsilon_{\mu\alpha\sigma} 
  k_2^\sigma(k_{2\nu} \eta_{\rho\beta} -k_{2\rho} \eta_{\beta\nu}) q_\lambda
\Big{)}\label{CS4}
\end{eqnarray}
\begin{eqnarray}
-\frac{\kappa}{4}\,\frac i8\,\Big{(}\!&&\!  
 \epsilon_{\nu\rho\sigma} \left( q^\sigma q_{\alpha} (k_{2\mu}- k_{1\mu}) \eta_{\lambda\beta} -
                          k_1^\sigma k_{1\alpha} (k_{2\lambda}+q_{\lambda}) \eta_{\mu\beta}
                          \right) \0\\\0 &&
+\epsilon_{\nu\beta\sigma} \left(q^\sigma q_{\lambda} (k_{1\mu} - k_{2\mu}) \eta_{\alpha\rho} -
                           k_2^\sigma k_{2 \lambda} (k_{1\alpha}+q_{\alpha}) \eta_{\mu\rho}
                          \right) \\ &&
+\epsilon_{\beta\rho\sigma}\left(k_1^\sigma k_{1\mu}( k_{2\lambda}+q_{\lambda}) \eta_{\alpha\nu} -
                           k_2^\sigma k_{2\mu}( k_{1\alpha}+q_{\alpha})     \eta_{\lambda\nu}
                           \right) \Big{)}\label{CS5}
\end{eqnarray}
\begin{eqnarray}
-\frac{\kappa}{4}\,\frac{i}{8}
\Big{(}&&\epsilon_{\sigma\lambda\alpha} 
( 
\eta_{\beta\mu} \eta_{\rho\nu} (k_1^\sigma k_1 \!\cdot\! k_2-k_2^\sigma k_2 \!\cdot\! q) + 
\eta_{\beta\nu} \eta_{\rho\mu} ( k_1^\sigma k_1 \!\cdot\! q-k_2^\sigma k_1 \!\cdot\! k_2)) \0\\\0
&& + 
\epsilon_{\sigma\mu\alpha} ( 
\eta_{\beta\lambda} \eta_{\nu\rho} ( q^\sigma q \!\cdot\! k_2 +k_2^\sigma k_1 \!\cdot\! k_{2})
+\eta_{\beta\rho} \eta_{\nu\lambda} (- q^\sigma k_{1} \!\cdot\! q + k_2^\sigma q \!\cdot\! k_2)) \\
&& + \epsilon_{\sigma\mu\lambda} 
(\eta_{\nu\beta} \eta_{\rho\alpha} ( q^\sigma q \!\cdot\! k_1  +k_1^\sigma k_1 \!\cdot\! k_{2})
+\eta_{\nu\alpha} \eta_{\rho\beta} (-q^\sigma  k_{2} \!\cdot\! q +  k_1^\sigma q \!\cdot\! k_1 ))\Big{)}
\label{CS6}
\end{eqnarray}
\begin{eqnarray}
\frac{\kappa}{4}\,\frac i{8}\Big{[}\!&&\!\epsilon_{\sigma\beta\nu} \eta_{\mu\rho} k_2^\sigma
\left(\eta_{\alpha\lambda} k_2^2 -k_{2\lambda}k_{2\alpha}\right) 
+\epsilon_{\sigma\beta\lambda} \eta_{\mu\rho} k_2^\sigma \left(\eta_{\alpha\nu}
k_2^2 -k_{2\nu}k_{2\alpha}\right) \0\\
&&+\epsilon_{\sigma\rho\nu} \eta_{\mu\beta} k_1^\sigma
\left(\eta_{\alpha\lambda} k_1^2 -k_{1\lambda}k_{1\alpha}\right) 
+\epsilon_{\sigma\rho\alpha} \eta_{\mu\beta} k_1^\sigma \left(\eta_{\lambda\nu}
k_1^2 -k_{1\nu}k_{1\lambda}\right) \0\\
&&-\epsilon_{\sigma\nu\beta} \eta_{\alpha\rho} \eta_{\mu\lambda} q^\sigma q^2- 
\epsilon_{\sigma\nu\rho} \eta_{\alpha\mu} \eta_{\beta\lambda} q^\sigma
q^2+\epsilon_{\sigma\mu\rho}\eta_{\alpha\lambda}q^\sigma q_\beta q_\nu+  
\epsilon_{\sigma\mu\alpha}\eta_{\beta\rho}q^\sigma q_\lambda q_\nu
\Big{]}\label{CS7}
\end{eqnarray}

These terms must be simmetrized under $\mu\leftrightarrow \nu, \lambda
\leftrightarrow
\rho,\alpha \leftrightarrow \beta$. They are expected to correspond to
odd-parity
3-point e.m. tensor correlator.

\subsection{The IR limit of the 3-point e.m. correlator}
\label{sec:3rdorderCS2}
 
The 0-th order term, after adding the cross contribution, is given (up to an
overall multiplicative factor of $\frac 1{128\cdot 32\pi}$) by
\be 
\tilde T^{(odd,IR)}_{\mu\nu\alpha\beta\lambda\rho}(k_1,k_2)= 
\frac 1{256\pi} \sum_{i=1}^4
\ET^{(i)}_{\mu\nu\lambda\rho\alpha\beta}(k_1,k_2)\label{ToddIR}
\ee
where
\begin{eqnarray}
\ET^{(1)}_{\mu\nu\lambda\rho\alpha\beta}(k_1,k_2)&=&-\epsilon_{\sigma\beta\nu}
k_2^\sigma
\left[ \frac 43 k_1\!\cdot\!k_2 \left( \eta_{\rho\lambda}\eta_{\alpha\mu}
+ \eta_{\rho\alpha}\eta_{\lambda\mu}+
\eta_{\rho\mu}\eta_{\lambda\alpha}\right)+\frac 43 q_\alpha k_{2\mu}
\eta_{\rho\lambda} \right.
- \frac 43 k_{1\alpha} k_{2\lambda} \eta_{\rho\mu}\nonumber\\
&& - \frac 23 \eta_{\lambda\mu}(q_\alpha k_{1\rho}+k_{1\alpha} q_\rho+
k_{1\alpha}k_{2\rho}) +\frac 23\eta_{\lambda\alpha} (2 q_\rho k_{1\mu} 
+ k_{1\rho}(k_1-k_2)_\mu) \nonumber\\
&&\left. + \frac 43 k_{1\mu}q_\lambda \eta_{\alpha\rho} +\frac 23 \eta_{\mu
\alpha}(2q_\rho q_\lambda+k_{1\rho} q_\lambda + q_\rho  k_{2\lambda} +
k_{2\rho}k_{2\lambda})\right]\nonumber\\
  &+&\frac 23 \epsilon_{\sigma\beta\nu} k_1^\sigma k_{2\rho}   \Big{[} 
(k_1-k_2)_\mu \eta_{\lambda\alpha}+ (q+k_{2} )_\lambda
\eta_{\mu\alpha}- (q+  k_{1})_\alpha\eta_{\lambda\mu}\Big{]}\label{3p-o-1}
\end{eqnarray}

\begin{eqnarray}
\ET^{(2)}_{\mu\nu\lambda\rho\alpha\beta}(k_1,k_2)&=& -\epsilon_{\sigma\rho\nu}
k_1^\sigma
\left[ \frac 43 k_1\!\cdot\!k_2 \left( \eta_{\beta\lambda}\eta_{\alpha\mu}
+ \eta_{\alpha\beta}\eta_{\lambda\mu}+
\eta_{\beta\mu}\eta_{\lambda\alpha}\right)+\frac 43 q_\lambda k_{1\mu}
\eta_{\beta\alpha} \right.
- \frac 43 k_{1\alpha} k_{2\lambda} \eta_{\beta\mu}\nonumber\\
&&  + \frac 23 \eta_{\lambda\mu}(2q_\alpha q_\beta+k_{1\alpha} q_\beta +
q_\alpha k_{2\beta}+ k_{1\alpha} k_{1\beta}) +\frac 23\eta_{\lambda\alpha} (2
q_\beta k_{2\mu} 
+ k_{2\beta}(k_2-k_1)_\mu) \nonumber\\
&&\left.+ \frac 43 k_{2\mu}q_\alpha \eta_{\lambda\beta} -\frac 23 \eta_{\mu
\alpha}(q_\beta k_{2\lambda} +k_{2\beta} q_\lambda +  
k_{2\lambda}k_{1\beta})\right]\nonumber\\
 &-&\frac 23 \epsilon_{\sigma\rho\nu} k_2^\sigma k_{1\beta}  \Big{[}  
(k_2-k_1)_\mu \eta_{\lambda\alpha}- (q+k_{2} )_\lambda
\eta_{\mu\alpha}+ (q+  k_{1})_\alpha\eta_{\lambda\mu}\Big{]}\label{3p-o-2}
\end{eqnarray}

\begin{eqnarray}
\ET^{(3)}_{\mu\nu\lambda\rho\alpha\beta}(k_1,k_2)&=& \epsilon_{\rho\beta\nu}
\left[\frac
{74}{15}  k_1\!\cdot\!k_2  (k_1-k_2)_\mu \eta_{\alpha\lambda}
-\frac 13 k_1\!\cdot\!k_2 (15 k_2+44 k_1)_\alpha  \eta_{\lambda\mu}\right.\label{3p-o-3}\\
&&+\frac 13  k_1\!\cdot\!k_2 (44 k_2+15 k_1)_\lambda  \eta_{\alpha\mu}-\frac {1}{15}
k_{1\alpha} k_{1\lambda}(11 k_{1}
+47k_{2})_\mu \nonumber\\
&&\left. +\frac {1}{15} k_{2\alpha} k_{2\lambda}(4 k_2+7k_1)_\mu+\frac 1{5}
k_{1\alpha} k_{2\lambda} (k_2-k_1)_\mu
+\frac {1}{15} k_{2\alpha}  k_{1\lambda}(37k_1+3k_2)_\mu \right]\nonumber
\end{eqnarray}

\begin{multline}\label{3p-o-4}
\ET^{(4)}_{\mu\nu\lambda\rho\alpha\beta} = -\eta_{\rho\nu}
\epsilon_{\sigma\beta\tau} k_1^\sigma k_2^\tau 
\left( \frac 23 \eta_{\mu\alpha}(k_1+2k_2)_\lambda +  \frac 23
\eta_{\lambda\alpha}(k_1-k_2)_\mu 
-  \frac 23 \eta_{\mu\lambda}(2k_1+k_2)_\alpha\right)\\
-\eta_{\beta\nu} \epsilon_{\sigma\rho\tau} k_2^\sigma k_1^\tau 
\left( -\frac 23 \eta_{\mu\alpha}(k_1+2k_2)_\lambda +  \frac 23
\eta_{\lambda\alpha}(k_2-k_1)_\mu 
+  \frac 23 \eta_{\mu\lambda}(2k_1+k_2)_\alpha\right).
\end{multline}
This must be simmetrized under $\mu\leftrightarrow \nu, \lambda \leftrightarrow
\rho,\alpha \leftrightarrow \beta$. The IR limit is entirely local.


\begin{thebibliography}{1}

\bibitem{Maldacena}
  J.~M.~Maldacena and G.~L.~Pimentel,
{\it On graviton non-Gaussianities during inflation,}
  JHEP {\bf 1109} (2011) 045, [arXiv:1104.2846 [hep-th]].

X.~O.~Camanho, J.~D.~Edelstein, J.~Maldacena and A.~Zhiboedov,
{\it Causality Constraints on Corrections to the Graviton Three-Point Coupling,}
  JHEP {\bf 1602} (2016) 020
   [arXiv:1407.5597 [hep-th]].

\bibitem{Huh} 
  Y.~Huh, P.~Strack and S.~Sachdev,
{\it Conserved current correlators of conformal field theories in 2+1 dimensions,}
  Phys.\ Rev.\ B {\bf 88}, 155109 (2013)
  [Phys.\ Rev.\ B {\bf 90}, no. 19, 199902 (2014)]
   [arXiv:1307.6863 [cond-mat.str-el]].

\bibitem{Vasiliev}
  S.~F.~Prokushkin and M.~A.~Vasiliev,
 {\it Higher spin gauge interactions for massive matter fields in 3-D AdS space-time,}
  Nucl.\ Phys.\ B {\bf 545} (1999) 385
 [hep-th/9806236].


\bibitem{Maldacena-Zhiboedov}
J.~Maldacena and A.~Zhiboedov,
{\it Constraining Conformal Field Theories with A Higher Spin Symmetry}
  J.\ Phys.\ A {\bf 46} (2013) 214011 [arXiv:1112.1016 [hep-th]].

J.~Maldacena and A.~Zhiboedov,
 {\it Constraining conformal field theories with a slightly broken higher spin symmetry,}
  Class.\ Quant.\ Grav.\  {\bf 30} (2013) 104003.
    [arXiv:1204.3882 [hep-th]].   

\bibitem{Closset}
Cyril Closset, Thomas~T. Dumitrescu, Guido Festuccia, Zohar Komargodski, and
  Nathan Seiberg, \emph{{Comments on Chern-Simons Contact Terms in Three
  Dimensions}}, JHEP \textbf{1209} (2012), 091.

\bibitem{giombi}  Simone Giombi, Shiroman Prakash and Xi Yin,
 \emph{A note on CFT correlators in Three Dimensions}, 
[arXiv:1104.4317[hep-th]].

\bibitem{GMPTWY} 
  S.~Giombi, S.~Minwalla, S.~Prakash, S.~P.~Trivedi, S.~R.~Wadia and X.~Yin,
  Eur.\ Phys.\ J.\ C {\bf 72}, 2112 (2012)
  [arXiv:1110.4386 [hep-th]].
 
\bibitem{Babu}
  K.~S.~Babu, A.~K.~Das and P.~Panigrahi,
  {\it Derivative Expansion and the Induced Chern-simons Term at Finite
Temperature in (2+1)-dimensions,}
  Phys.\ Rev.\ D {\bf 36} (1987) 3725.

\bibitem{Dunne}
  G.~V.~Dunne,
  {\it Aspects of Chern-Simons theory,}
  hep-th/9902115.
  
\bibitem{Gama}
  F.~S.~Gama, J.~R.~Nascimento and A.~Y.~Petrov,
  {\it Derivative expansion and the induced Chern-Simons term in N=1, d=3
superspace,}
  arXiv:1511.05471 [hep-th].
  
\bibitem{bekaert}  X.~Bekaert, E.~Joung and J.~Mourad,
{\it Effective action in a higher-spin background,}
JHEP {\bf 1102}, 048 (2011), [arXiv:1012.2103 [hep-th]].

\bibitem{Witten}
E. Witten, {\it Anomalies Revisited}, Lecture At Strings 2015, ICTS-TITR,
Bangalore, June 22, 2015.

E. Witten, {\it Fermion Path Integrals And Topological Phases},
arXiv:hep-th/1508.04715.

\bibitem{BL} 
  L.~Bonora and B.~L.~de Souza,
  {\it Pure contact term correlators in CFT,}
  arXiv:1511.06635 [hep-th].

\bibitem{BGL} 
  L.~Bonora, S.~Giaccari and B.~Lima de Souza,
  {\it Trace anomalies in chiral theories revisited,}
  JHEP {\textbf 1407}, 117 (2014)
  [arXiv:1403.2606 [hep-th]].	
 	
\bibitem{BGLBled13} 
  L.~Bonora, S.~Giaccari and B.~L.~D.~Souza,
  {\it Revisiting Trace Anomalies in Chiral Theories,}
  Springer Proc.\ Math.\ Stat.\  {\bf 111}, 3 (2014)

\bibitem{BDL}  L.~Bonora, A.~D.~Pereira and B.~L.~de Souza,
 {\it Regularization of energy-momentum tensor correlators and parity-odd
terms,}
  JHEP {\bf 1506}, 024 (2015)
  [arXiv:1503.03326 [hep-th]].

\bibitem{Vuorio}
  I.~Vuorio,
  {\it Parity Violation and the Effective Gravitational Action in
Three-dimensions,}
  Phys.\ Lett.\ B {\bf 175} (1986) 176.
 
\bibitem{pope} C.~N.~Pope and P.~K.~Townsend
{\it Conformal higher spins in (2+1) dimensions}
Phys.Lett. {\bf B225} (1989) 245. 

\bibitem{BoosDavy} E. E. Boos and Andrei I. Davydychev. {\it A Method of
evaluating massive Feynman integrals}, Theor.
Math. Phys., {\bf 89}  (1991) 1052 [Teor. Mat. Fiz.89,56(1991)].

Andrei I. Davydychev.{\it A Simple formula for reducing Feynman diagrams to
scalar integrals}, Phys. Lett.,
{\bf B263} (1991) 107

Andrei I. Davydychev. {\it Recursive algorithm of evaluating vertex type Feynman
integrals}, J. Phys.,
A25 (1992) 5587.
\bibitem{solvay2004} {\it Higher-Spin Gauge Theories}, Proceedings of the First
Solvay Workshop, held in Brussels on May 12-14, 2004, eds. R. Argurio, G. Barnich, G. Bonelli and M. Grigoriev (Int. Solvay Institutes, 2006).
 
\bibitem{sorokin} D.~ Sorokin, AIP Conf. Proc.767 (2005) 172
[hep-th/0405069]; D. Francia, A. Sagnotti, J. Phys. Conf. Ser. 33 (2006) 57 [hep-th/0601199]; 
A. Fotopoulos, M. Tsulaia, Int. J. Mod. Phys. {\bf A24} (2009) 1 [arXiv:0805.1346]; 
C. Iazeolla, arXiv:0807.0406; A. Campoleoni, Riv.Nuovo Cim. {\bf 033} (2010) 123 [arXiv:0910.3155]; 
A. Sagnotti, arXiv:1002.33 88; D. Francia, Prog. Theor. Phys. Suppl. {\bf 188} (2011) 94 [arXiv:1103.0683].

\bibitem{campoleoni} A.~Campoleoni, {\it Higher Spins in D = 2 + 1,}
  Subnucl.\ Ser.\  {\bf 49} (2013) 385.
  [arXiv:1110.5841 [hep-th]].

\bibitem{blencowe}  M.~ P.~ Blencowe, {\it A consistent  interacting 
massless higher-spin field  theory in D=2+1}
Class.Quant.Grav. {\bf 6} (1989)  443

\bibitem{WittenCS} E.~Witten, {\it 2+1 dimensional gravity as an exact soluble system.}
Nucl.Phys. {\bf B 311} (1988) 46.

\bibitem{deWit:1979sib}
  B.~de Wit and D.~Z.~Freedman,
  {\it Systematics of Higher Spin Gauge Fields,}
  Phys.\ Rev.\ D {\bf 21} (1980) 358.

\bibitem{Damour:1987vm}
  T.~Damour and S.~Deser,
  {\it 'Geometry' of Spin 3 Gauge Theories},
  Annales Poincare Phys.\ Theor.\  {\bf 47} (1987) 277.

 \bibitem{Parker:2009}
L.~Parker and D.~Toms, \emph{{Quantum Field Theory in Curved Spacetime}},
  (2009).

\end{thebibliography}
 \end{document}